\documentclass[12pt]{article}

\newlength{\minitwocolumn}
\usepackage{amsmath}
\usepackage{amssymb,amscd}
\usepackage{bm}
\usepackage{amsbsy} 

\newcommand{\nom}{\nonumber}
\newcommand{\beq}{\begin{equation}}
\newcommand{\eeq}{\end{equation}}
\newcommand{\bea}{\begin{eqnarray*}}
\newcommand{\eea}{\end{eqnarray*}}
\newcommand{\beqa}{\begin{eqnarray}}
\newcommand{\eeqa}{\end{eqnarray}}

\newcommand{\bx}{\boldsymbol{x}}
\newcommand{\bq}{\boldsymbol{q}}
\newcommand{\bp}{\boldsymbol{p}}

\newcommand{\bJ}{\boldsymbol{J}}

\newcommand{\veta}{\boldsymbol{\eta}}
\newcommand{\vxi}{\boldsymbol{\xi}}
\newcommand{\vchi}{{\boldsymbol{\chi}}}
\newcommand{\vThe}{{\boldsymbol{\Theta}}}
\newcommand{\del}{\partial}
\newcommand{\ti}{\tilde}

\def\bR{\mbox{\boldmath $R$}}

\newcommand{\calL}{{\cal L}}
\newcommand{\calM}{{\cal M}}

\def\bOmega{\mbox{$\Omega$}}
\def\bbOmega{\mbox{\boldmath $\Omega$}}

\newcommand{\Map}{{\rm Map}}
\newcommand{\ev}{{\Phi}}

\newcommand{\pbv}[2]{{\{{{#1},{#2}}\}}}
\newcommand{\sbv}[2]{{\{{{#1},{#2}}\}}}
\newcommand{\courant}[2]{{[{{#1},{#2}}]}}

\newtheorem{theorem}{Theorem}[section]

\newtheorem{definition}[theorem]{Definition}

\newtheorem{example}[theorem]{Example}

\newcommand{\qed}{\nobreak \ifvmode \relax \else
      \ifdim\lastskip<1.5em \hskip-\lastskip
      \hskip1.5em plus0em minus0.5em \fi \nobreak
      \vrule height0.75em width0.5em depth0.25em\fi}

\newcommand{\be}{\boldsymbol{e}}
\newcommand{\floor}[1]{{\lfloor #1 \rfloor}}
\newcommand{\bracket}[2]{\langle #1\,,#2\rangle}
\newcommand{\hatj}{{\widehat{j}}}

\begin{document}

\baselineskip 0.7cm

\begin{titlepage}
\vspace{4\baselineskip}
\begin{flushright}
MISC-2011-14
\end{flushright}
\vspace{0.6cm}
\begin{center}
{\Large\bf  
Current Algebras and QP Manifolds}
\end{center}
\vspace{1cm}
\begin{center}
{Noriaki IKEDA
\footnote{E-mail: nikeda@se.ritsumei.ac.jp, 
ikeda@yukawa.kyoto-u.ac.jp}}
and
{Kozo KOIZUMI
\footnote{E-mail: kkoizumi@cc.kyoto-su.ac.jp}}
\end{center}
\vspace{0.2cm}
\begin{center}
{
\it Maskawa Institute for Science and Culture,
Kyoto Sangyo University \\ 
Kyoto 603-8555, Japan}
\medskip
\vskip 10mm
\end{center}
\vskip 10mm
\begin{abstract}
Generalized current algebras introduced by Alekseev and Strobl
in two dimensions are reconstructed by a graded manifold
and a graded Poisson brackets. 
We generalize their current algebras to higher dimensions.
QP manifolds provide the unified structures of current algebras in any dimension.
Current algebras give rise to structures of Leibniz/Loday algebroids,
which are characterized by QP structures.
Especially, in three dimensions, a current algebra has a structure of 
a Lie algebroid up to homotopy introduced by Uchino and one of the authors
which has a bracket of a generalization of the Courant-Dorfman bracket.
Anomaly cancellation conditions are reinterpreted as
generalizations of the Dirac structure.
\end{abstract}
\vspace{0.1cm}
\vskip 0.8cm

\end{titlepage}

\section{Introduction}
\noindent
Alekseev and Strobl \cite{Alekseev:2004np} have generalized 
a current algebra in two dimensions to a target space $TM \oplus T^*M$,
which is described by the Courant bracket,
where $M$ is a manifold in $d$ dimensions.
The condition that currents close and does not have anomalies
is geometrically characterized by the Dirac structure on $TM \oplus T^*M$.
Since the current algebra naturally contains fluxes, this is related to the string theory
with a flux background. Generalizations of their current algebras
to higher dimensions \cite{Bonelli:2005ti}, or more general currents in two dimensions
\cite{Ekstrand:2009qz} have been constructed.

The canonical commutation relations of canonical conjugates
can be reformulated in terms of the supergeometry.
A Poisson bracket is constructed from a Schouten-Nijenhuis bracket 
and a Poisson bivector field \cite{vaisman}. We generalize this formulation 
to current algebras.

The algebra to be closed under the Courant bracket
is the Courant algebroid \cite{Courant}, \cite{LWX}.
The Courant algebroid has a supermanifold 
construction \cite{Roy01} by the derived bracket \cite{KosmannSchwarzbach:2003en}.
This construction is closely related to a QP structure of 
a topological field theory \cite{Schwarz:1992nx} in three dimensions.
In fact, the supermanifold construction of the Courant bracket has derived 
a topological sigma model in three dimensions, which is called
the Courant sigma model \cite{Ikeda:2002wh}, \cite{Roytenberg:2006qz}.
This is a direct application of a AKSZ construction 
\cite{Alexandrov:1995kv}, \cite{Cattaneo:2001ys}, \cite{Roytenberg:2006qz}
in a topological field theory to three dimensions.
The supergeometry is a key idea again.

In this paper, 
current algebras described in Alekseev and Strobl
are reconstructed in terms of a QP structure.
This reconstruction proposes the unified structure of 
current algebras in an arbitrary dimension.
Moreover current algebras 
\cite{Bonelli:2005ti}
in a $n$ dimensions worldvolume are
consistently generalized 
in order to contain more general dynamical systems.
We point out that current 'algebras' have structures 
such as the Lie algebroids  \cite{Mackenzie}, the Courant algebroid
and their generalizations, 
containing the Loday algebroids.
Especially, current algebra in three dimensions 
has a structure of 
a Lie algebroid up to homotopy \cite{Ikeda:2010vz},
more generally, the H-twisted Lie algebroid \cite{Grutzmann}.
Anomaly cancellation conditions are clarified
in terms of QP manifolds of degree $n$
and gives rise to
a generalization of the Dirac structure.

The paper is organized as follows. 
In section 2, a supermanifold construction of current algebras 
in one dimension is considered. In section 3, the paper of Alekseev 
and Strobl is reviewed. In section 4, mathematics related to this article 
is explained. In section 5, the current algebras in two dimensions are reconstructed 
from a QP manifold. In section 6, the generalized 
current algebras in three dimensions are introduced 
and reinterpreted from a QP manifold.
In section 7, the generalized current algebras in $n$ dimensions are discussed.
Section 8 is conclusions and discussion.

\section{Current Algebras in One Dimension
}
\noindent
First we consider one dimensional worldline case
as a typical and well known example of our ideas.
Let us consider a space $M$ of $d$ dimensions
defined on one dimensional worldline $X_1 = \bR$, 
which possesses the canonical conjugates ($x^I, p_J$)
on a phase space $T^*M$,
where the Poisson brackets are given by
\beq
    \{x^I,x^J\}_{P.B.}=0,\quad 
    \{x^I,p_J\}_{P.B.}=\delta^I{}_J,\quad 
    \{p_I,p_J\}_{P.B.}=0,
\eeq
where $I, J, \cdots $ are indices on $M$.
Introducing a gauge potential $A_I(x)$ on $M$, 
the canonical momentum $p_I$ is shifted as $p_I\to p_I+A_I$. 
The Poisson brackets of canonical conjugates are twisted 
by a closed 2-form $H_{IJ}=\del_I A_J-\del_J A_I$ 
on the phase space as follows:
\beq
    \{x^I,x^J\}_{P.B.}=0,\quad 
    \{x^I,p_J\}_{P.B.}=\delta^I{}_J,\quad 
    \{p_I,p_J\}_{P.B.}=-H_{IJ}.
\label{1dpoisson}
\eeq
A 'current' on the phase space is a function
$F(x,p)$  which does not depend on the coordinate $t$ on $X_1$
explicitly, that is, $\partial_t F(x, p) =0$.
Let $F(x,p)$ and $G(x,p)$ be currents on the phase space.
The Poisson bracket between 
two currents gives us a new current:
\beq
    \{F(x,p),G(x,p)\}_{P.B.}=\frac{\del F}{\del x^I}\frac{\del G}{\del p_I}
    -\frac{\del F}{\del p_I}\frac{\del G}{\del x^I}+H_{IJ}
    \frac{\del F}{\del p_I}\frac{\del G}{\del p_J} \equiv K(x,p),
\eeq
from Eq. (\ref{1dpoisson}).

This structure is reformulated by an odd Poisson bracket, 
which is called the Schouten-Nijenhuis bracket.
Let us consider the exterior algebra $\wedge^{\bullet} T(T^*M)$, 
of which sections are identified to functions on a supermanifold
$T^*[1](T^*M)$, 
$\Gamma \wedge^{\bullet} T(T^*M) = C^{\infty}(T^*[1](T^*M))$
\footnote{Recent reviews about a supermanifold and a supergeometry
are \cite{Kotov:2010wr}, \cite{Cattaneo:2010re} and \cite{Qiu:2011qr}.}.
An odd symplectic form $\Omega$ is induced 
from a natural symplectic structure on a (double) cotangent bundle 
$T^*(T^*M)$, 
which is 
\beq
    \Omega=
    \ \delta \bx^{\ti I}
\wedge \delta \vxi_{\ti I}
=    \delta x^{I} \wedge \delta \xi_{I}
+ \delta p_{I} \wedge \delta \eta^{I},
\eeq
where
$\bx^{\ti I} = (x^I, p_I)$ is a Darboux coordinate on 
$T^*M$ and $\vxi_{\ti I} = (\xi_I, \eta^I)$ is an odd local coordinate 
of the fiber of $T^*[1](T^*M)$.
Degrees are assigned to each coordinate.
$\bx^{\ti I} = (x^I, p_I)$ has degree $0$ and
$\vxi_{\ti I} = (\xi_I, \eta^I)$ has degree $1$.
The odd Poisson brackets on the canonical quantities are 
\beq
     \{\bx^{\ti I}, \bx^{\ti J}\}=0,\quad
     \{\vxi_{\ti I}, \vxi_{\ti J}\}=0,\quad
     \{\bx^{\ti I}, \vxi_{\ti J}\}=\delta^{\ti I}{}_{\ti J}.
\eeq

Now let us require a degree $2$ function 
$\Theta \in C^{\infty}(T^*[1](T^*M))$ 
such that 
$
\{\Theta,\Theta\}=0.
$
A general solution of $\Theta$ is
\beq
    \Theta=\frac{1}{2}f^{{\ti I} {\ti J} }(\bx)\vxi_{\ti I}\vxi_{\ti J},
\eeq
where $f^{{\ti I} {\ti J} }(\bx)$ is skewsymmetric
and $\frac{\partial f^{{\ti I} {\ti J} }(\bx)}{\partial \bx^{\ti L}}
f^{{\ti L} {\ti K}}(\bx) + ({\ti I}{\ti J}{\ti K} \ \mbox{cyclic})=0$.
Since no background structure except for $H_{IJ}$ 
is assumed on $M$, the only solution is 
\beq
    f^{{\ti I}{\ti J}}(\bx)=
    \left(\begin{array}{cc}
            0&-\delta^{I}{}_{J}\\
            \delta^{J}{}_{I}&H_{I J}(\bx)\end{array}\right).
\eeq
Here we set $\ti I=(1,\cdots,D,D+1,\cdots,2D)$ and
$ \bx^{\ti I}= (\bx^I, \bx^{D+I}) = (x^I, p_I)$.
Under these settings, the original Poisson bracket
is reconstructed by the derived bracket:
\beq
\pbv{F(x, p)}{G(x,p)}_{P.B.} = \sbv{\sbv{F(\bx)}{\Theta}}{G(\bx)}.
\eeq
In fact, the Poisson brackets on the canonical conjugates are derived as
\beq
    \{\{x^{I},\Theta\}, x^{J}\}=0,\quad
    \{\{x^{I},\Theta\}, p_{J}\}={\delta^I}_J,\quad
    \{\{p_{I},\Theta\}, p_{J}\}=-H_{IJ}.
\eeq
The construction here is known 
as a Poisson bracket 
from a Schouten-Nijenhuis bracket and a Poisson bivector field
in the Poisson geometry \cite{vaisman}.
A Schouten-Nijenhuis bracket $\sbv{-}{-}$ is
an odd Poisson bracket and a Poisson bivector field is $\Theta$.
The Poisson structure is associated to
a structure with a Lie algebroid on $T(T^*M)$.


\section{Current Algebras in Two Dimensions}
\noindent
In this section, the construction of 
current algebras by Alekseev and Strobl 
\cite{Alekseev:2004np}
in two dimensional spacetime is reviewed.
Let us consider a two dimensional worldsheet $X_2 = S^1 \times \bR$.
The phase space is the cotangent bundle $T^*LM$ of the loop space 
$LM= \Map(S^1, T^*M)$.
Let $x^I(\sigma)$ be a local coordinate and 
$p_I(\sigma)$ be a canonical conjugate, 
where $\sigma$ is a local coordinate on $S^1$.
The symplectic structure can be written as follows:
\beq
    \omega=\int_{S^1}d\sigma \ \delta x^I\wedge \delta p_I.
\label{2dsymplectic1}
\eeq
The Poisson bracket on the canonical quantities is
\beq
    \{x^I, x^J\}_{P.B.}=0,\quad 
    \{x^I, p_J\}_{P.B.}=\delta^I{}_J \delta(\sigma - \sigma^{\prime}),\quad 
    \{p_I, p_J\}_{P.B.}=0.
\label{2dpoisson1}
\eeq
More generally, 
(\ref{2dsymplectic1}) can be twisted by a closed $3$-form $H$ as
\beq
    \omega=
\int_{S^1}d\sigma \ \delta x^I\wedge \delta p_I + 
\frac{1}{2} \int_{S^1} d\sigma \ H_{IJK} 
\partial_{\sigma} x^I  \delta x^J \wedge \delta x^K.
\label{2dsymplectic}
\eeq
Then the Poisson bracket is modified to
\beq
    \{x^I,x^J\}_{P.B.}=0, \quad 
    \{x^I,p_J\}_{P.B.}=\delta^I{}_J \delta(\sigma - \sigma^{\prime}),\quad 
    \{p_I,p_J\}_{P.B.}= - H_{IJK} \partial_{\sigma} x^K 
\delta(\sigma - \sigma^{\prime}).
\label{2dpoisson}
\eeq

A generalization of a current algebra 
to a target space $TM \oplus T^*M$
has considered:
\beq
    J_{0(f)}(\sigma)=f(x(\sigma)), \quad
    J_{1(u,\alpha)}(\sigma)
     =\alpha_I(x(\sigma)) \partial_{\sigma} x^I(\sigma) 
+ u^I(x(\sigma))p_I(\sigma),
\label{2dcurrent}
\eeq
where $f(x(\sigma))$ is a function,
$\alpha_I(x) d x^{I}$ is a $1$-form and 
$u=u^I(x) \partial_I$ is a vector field
on the target space.
$J_{0(f)}$ is a current of mass dimension zero and 
$J_{1(u,\alpha)}$ is one of mass dimension one.
These forms contain 
current Kac-Moody algebras on the WZW model, currents of 
the Poisson sigma model and 
a Killing vector field with a $3$-form on the target space
as special cases.

The canonical commutation relations (\ref{2dpoisson}) 
derive commutation relations of current algebras
(\ref{2dcurrent}):
\beqa
    \{J_{0(f)}(\sigma),J_{0(f^\prime)}(\sigma^\prime)\}_{P.B.}&=&0, 
\nonumber \\
    \{J_{1(u,\alpha)}(\sigma),J_{0(f^\prime)}(\sigma^\prime)\}_{P.B.}
&=& -u^{I}\frac{\del f^\prime}{\del x^I}(x(\sigma)) 
\delta(\sigma-\sigma^\prime),
\nonumber \\
\{J_{1(u,\alpha)}(\sigma),J_{1(u^\prime,\alpha^\prime)}(\sigma^\prime)\}_{P.B.}
  &=&
-J_{1(\courant{(u,\alpha)}{(u^\prime,\alpha^\prime)})}(\sigma)
\delta(\sigma-\sigma^\prime)
\nom\\
  &&+
\langle (u, \alpha), (u^\prime,\alpha^\prime ) \rangle
(\sigma^\prime)
\partial_{\sigma} \delta(\sigma-\sigma^\prime),
\label{2dcurrentalgebra}
\eeqa
where
\beq
    \courant{(u,\alpha)}{(u^\prime,\alpha^\prime)}
       =([u,u^\prime], L_u\alpha^\prime-L_{u^\prime}\alpha
        +d (i_{u^\prime} \alpha) + H(u,u^\prime, \cdot\ )),
\eeq
is the Courant-Dorfman bracket on $TM \oplus T^*M$
and 
$\langle (u, \alpha), (u^\prime,\alpha^\prime ) \rangle
= i_{u^{\prime}} \alpha + i_{u} \alpha^{\prime}
$ is a symmetric scalar product on $TM \oplus T^*M$
\cite{Courant}, \cite{LWX}.

From Eq.~(\ref{2dcurrentalgebra}),
the anomaly cancellation condition is 
$\langle (u, \alpha), (u^\prime,\alpha^\prime ) \rangle =0$.
The current algebra closes if this condition is satisfied.
This is satisfied 
on the Dirac structure on $M$.
The Dirac structure is a maximally isotropic subbundle of $TM \oplus T^*M$, 
whose sections are closed under the Courant-Dorfman bracket.

\section{QP Manifold}
\noindent
In this section, mathematical structures
which appear in this paper,
such as a QP manifold and an algebroid
are prepared.

A nonnegatively graded manifold $\cal M$, called a N-manifold,
is defined as a ringed space 
with a structure sheaf 
of nonnegatively graded commutative algebra over an ordinary smooth manifold 
$M$. 
Grading is called \textbf{degree}.

A N-manifold equipped with 
a graded symplectic structure (\textbf{P-structure}) 
$\bOmega$ of degree $n$ is called a P-manifold of degree $n$,
 $({\cal M},\bOmega)$.
The graded Poisson bracket on $C^\infty ({\cal M})$ is defined from the graded symplectic
structure $\bOmega$ on ${\cal M}$ as 
$    \{f,g\} = (-1)^{|f|+1} i_{X_f} i_{X_g}\bOmega,$
where a Hamiltonian vector field $X_f$ is defined by the equation 
$\{f,g\} = X_f g$, 
for $f,g \in C^\infty({\cal M})$. 
\begin{definition}
Let $({\cal M}, \bOmega)$ be a $P$-manifold of degree $n$
and $Q$ be a differential of degree $+1$ with $Q^2=0$ on $\calM$.
$Q$ is called a \textbf{Q-structure}.
A triple $(\calM, \bOmega, Q)$ is called
a \textbf{QP-manifold} of degree $n$ 
and its structure is called a \textbf{QP structure},
if $\bOmega$ and $Q$ are compatible,
that is, $\calL_Q \bOmega =0$ {\rm \cite{Schwarz:1992nx}}.
\end{definition}
$Q$ is also called a homological vector field. 
A Hamiltonian $\Theta\in C^{\infty}(\calM)$ 
of $Q$ with respect to the graded Poisson bracket $\{-,-\}$
satisfies
\beq
Q=\{\Theta,-\},
\eeq
and has degree $n+1$.
The differential condition, $Q^2=0$, implies that
$\Theta$ is a solution of the \textbf{classical master equation},
\begin{equation}
\{\Theta,\Theta\}=0.
\label{cmaseq}
\end{equation}
A QP manifold $(\calM, \bOmega, Q)$ is also denoted by 
$(\calM, \bOmega, \Theta)$.

$T^*[1](T^*M)$ in section 2 is a graded manifold of degree $1$.
An odd symplectic form $\bOmega$ is a P-structure of degree $1$
and a Poisson bivector field $\Theta$ is a Q-structure.
Thus a current algebra in one dimension has
a structure of a QP-manifold of degree $1$.

\begin{definition}
A vector bundle $E = (E, \rho, [-,-])$ is called \textbf{an algebroid}
if there is a bracket product 
$[e_{1},e_{2}]$, where $e_{1},e_{2}\in\Gamma E$, 
and a bundle map $\rho:E\to TM$ which is called an anchor map,
satisfying the conditions below:
\begin{eqnarray}
&& \rho[e_{1},e_{2}]=[\rho(e_{1}),\rho(e_{2})], 
\label{algebroid1} \\
&& [e_{1},fe_{2}]=f[e_{1},e_{2}]+\rho(e_{1})(f)e_{2},
\label{algebroid2}
\end{eqnarray}
where the bracket $[\rho(e_{1}),\rho(e_{2})]$
is the usual Lie bracket on $\Gamma TM$.
\end{definition}
A Loday algebroid is an algebroid version of a Loday(Leibniz) algebra
\cite{LP}, \cite{Loday}.
\begin{definition}
An algebroid $E = (E, \rho, [-,-])$ is called \textbf{a Loday algebroid}
if there is a bracket product 
$[e_{1},e_{2}]$ satisfying the Leibniz identity:
\begin{eqnarray}\label{defcou}
\label{loday1}
[e_1 ,[e_2 , e_3]]&=&[[e_1, e_2], e_3]+[e_2 ,[e_1, e_3]],
\end{eqnarray}
where $e_{1},e_{2}, e_3 \in\Gamma E$, 
A Loday algebroid is also called a Leibniz algebroid.
\end{definition}
Correspondence of a Loday algebroid 
with a homological vector field on a supermanifold is 
discussed in \cite{GKP}.
The following theorem has appeared in \cite{Kotov:2010wr}.
\begin{theorem}\label{QPloday}
Let $n > 1$. 
Functions of degree $n-1$ on a QP manifold can be identifies 
as sections of a vector bundle $E$.
The QP-structure induces a Loday algebroid structure on $E$.
\end{theorem}
%
Let $\bx$ be an element of degree $0$ and 
$\be^{(n-1)}$ be the element of degree $n-1$.
If we define
\begin{eqnarray}
[e_1, e_2]
&=& \sbv{\sbv{\be_1^{(n-1)}}{\Theta}}{\be_2^{(n-1)}}, \\
\rho(e _1) F(x)
&=& \sbv{\sbv{\be^{(n-1)}}{\Theta}}{F(\bx)},
\end{eqnarray}
$[-,-]$ and $\rho$ satisfy the relations of a Loday algebroid
(\ref{algebroid1}), (\ref{algebroid2})
and (\ref{loday1}).


Examples of a QP manifold of degree $n$ are listed.
\begin{example}\label{LieAlgebra}
{\rm 
Let $\mathfrak{g}$ be a Lie algebra with a Lie bracket $[-,-]$.
Then $T^*[n]\mathfrak{g}[1]$ is a QP manifold of degree $n$.
A natural P-structure $\bOmega$ is induced from 
the canonical symplectic structure 
on $T^*\mathfrak{g}$,
constructed from the canonical pairing of $\mathfrak{g}$ 
and $\mathfrak{g}^*$, $\langle -, - \rangle$.
Define $\Theta = \frac{1}{2} \langle \bp, [\bq, \bq] \rangle$, 
where $\bq \in \mathfrak{g}[1]$ and $\bp \in \mathfrak{g}^*[n]$.
Since $\sbv{\Theta}{\Theta}=0$ from a Lie algebra structure, 
$\Theta$ defines a Q-structure.
If we take a structure constant $f^A{}_{BC}$ of Lie algebra,
$\Theta = \frac{1}{2} f^A{}_{BC} \bp_A \bq^B \bq^C$.
}
\end{example}

\begin{example}\label{pmanifold}
{\rm 
Let $n=1$. Then $\calM$ is canonically 
$
\calM=T^{*}[1]M
$
and
a Poisson bracket $\sbv{-}{-}$ 
in the P-structure is a Schouten-Nijenhuis 
bracket. A Q-structure $\Theta$ has degree $2$
and $Q^2=0$ is that $\Theta$ is a Poisson bivector field.
Thus a QP manifold of degree $1$ is a Poisson manifold on $M$.
}
\end{example}

\begin{example}\label{Calgebroid}
{\rm 
Let $n=2$. 
A P-structure 
$\bOmega$ is an even form of degree $2$. 
A Q-structure $\Theta$ has degree $3$
and $Q^2=0$ defines a Courant algebroid structure on a vector bundle $E$.
A QP manifold of degree $2$ is a Courant algebroid \cite{Roy01}.
}
\end{example}
The Dirac structure $L$ is a 
maximally isotropic subbundle of the Courant algebroid $E$, 
whose sections are closed under the Courant-Dorfman bracket.
The symmetric scalar product $\langle -, - \rangle$ 
corresponds to
the Q-structure $\sbv{-}{-}$ of a QP manifold construction
of the Courant algebroid.
If  we identify functions on a QP manifold 
to the sections of a vector bundle $E$, 
the sections of the Dirac structure $\Gamma L$ 
are commutative under the P-structure $\sbv{-}{-}$ 
and closed under the derived bracket $\sbv{\sbv{-}{\Theta}}{-}$.

\begin{example}\label{Talgebroid}
{\rm 
Let $n=3$. 
One of examples of a N-manifold is 
$
\calM:=T^{*}[3]E[1]
$.
A P-structure $\bOmega$ is an odd form of degree $3$.
A Q-structure $Q^2=0$ defines a Lie algebroid up to homotopy 
(the splittable H-twisted Lie algebroid) on $E$ \cite{Ikeda:2010vz}.
A general nonsplittable algebroid is the H-twisted Lie algebroid
\cite{Grutzmann}.
}
\end{example}

\begin{example}\label{liealoid}
{\rm Let $E$ be a vector bundle on $M$ and
$
\calM = T^{*}[n]E[1]$.
If a QP structure is defined on $\calM = T^{*}[n]E[1]$
and $n\ge 4$, 
$E$ becomes a Lie algebroid.
(A Lie algebroid is a Loday algebroid which bracket 
$[-,-]$ is skewsymmetric.)
$\Gamma E\oplus\wedge^{n-1}E^{*}$ is a subalgebroid.
A QP structure 
on $T^*[n]E[1]$
induces the Courant-Dorfman bracket on $E\oplus\wedge^{n-1}E^{*}$
by the derived bracket $[-,-] = \sbv{\sbv{-}{\Theta}}{-}$
\cite{Zambon:2010ka}\cite{Ikeda:2012pv},
which has the following form,
\beq
[u+\alpha,v+\beta]=[u,v]+L_{u}\beta-i_{v}d\alpha+H(u,v),
\eeq
where
$u,v\in\Gamma E$, $\alpha,\beta\in\Gamma\wedge^{n-1}E^{*}$
and $H$ is a closed $(n+1)$-form on $E$.
We refer the reader to \cite{Hagiwara}, \cite{Wade} and \cite{BS} 
for detailed studies of the bracket of this type.
QP descriptions of higher Courant-Dorfman brackets are discussed in
\cite{Zambon}.
}
\end{example}

\section{QP Structures of Current Algebras in Two Dimensions}
\noindent
The construction of the mechanics by the
odd Poisson (Schouten-Nijenhuis) 
bracket in section 2 is generalized
to current algebras in two dimensions.
Let us consider the super extension of the space direction of 
the worldsheet $S^1$, $T[1]S^1$,
which has a local coordinate $(\sigma, \theta)$.
Moreover we consider the graded extension 
of degree $2$ of the target space,
$\calM =T^*[2](T^*[1]M)$
and the space of smooth map from $T[1]S^1$ to $T^*[2](T^*[1]M)$,
$\Map(T[1]S^1, T^*[2](T^*[1]M))$.
A QP structure of degree $2$ is introduced on 
$T^*[2](T^*[1]M)$.
Let $\Omega$ be a P-structure and 
$\Theta$ be a Q-structure on $T^*[2](T^*[1]M)$

$x^I(\sigma)$ is extended to a superfield 
$\bx^I(\sigma,\theta) = x^I(\sigma) + \theta x^{(1)I}(\sigma)$,
which is a smooth map from $T[1]S^1$ to $M$.
The canonical conjugate $p_I$ is extended to 
an odd superfield of degree $1$, 
$\bp_I(\sigma,\theta) = p^{(0)}_{I}(\sigma) + \theta p_I(\sigma)$,
which is a section of 
$T^*[1]S^1 \otimes \bx^*(T^*[1]M)$,
where $p^{(0)}_{I}$ is an odd auxiliary field of degree $1$,
Moreover Let us introduce 
'canonical conjugates' of 
$\bp_I$ and $\bx^I$, respectively.
A superfield
$\veta^I(\sigma,\theta) = \eta^{(0)I}(\sigma) + \theta \eta^{(1)I}(\sigma)$ 
of degree $1$ is a section of 
$T^*[1]S^1 \otimes \bx^*(T[1]M)$ and 
$\vxi_I(\sigma,\theta) = \xi^{(0)}_I(\sigma) + \theta \xi^{(1)}_I(\sigma)
$ of degree $2$ is a section of $T^*[1]S^1 \otimes \bx^*(T^*[2]M)$.
\medskip\\
\indent
The Courant algebroid structure on $TM \oplus T^*M$
is mapped by the canonical shifting and embedding 
called the minimal symplectic realization 
$j: TM \oplus T^*M \longrightarrow T^*[2]T^*[1]M$ such that
$j: (x^I, \frac{\partial}{\partial x^I}, 0, d x^I)
 \longmapsto (x^I, p_I, \xi_I, \eta^I)$.
$\frac{\partial}{\partial x^I}
 \longmapsto p_I$ is defined by 
a natural pairing of $TM$ and $T^*M$,
$\bracket{u^I \frac{\partial}{\partial x^I}}{p_I d x^I}
\mapsto {u^I}{p_I}$.
$j$
induces a map 
$\hatj: (x^I, p_I, 0, d x^I)
 \longmapsto (\bx^I, \bp_I, \vxi_I, \veta^I)$
for superfields.

A graded symplectic form 
$\bbOmega$ is  defined from $\Omega$ as
\beq
    \bbOmega= \int_{T[1]S^1} d \sigma d \theta \ \Phi^* \Omega
= 
\int_{T[1]S^1} d \sigma d \theta
    \ (\delta \bx^{I} \wedge\delta \vxi_{I}
    + \delta \bp_{I} \wedge\delta \veta^{I}),
\eeq
where $\Phi \in \Map(T[1]S^1, T^*[2](T^*[1]M))$.
Then the commutation relations are obtained as
\beqa
    &&\{\bp_J(\sigma,\theta), \veta^I(\sigma^\prime,\theta^\prime)\}
    =\{\veta^I(\sigma,\theta),\bp_J(\sigma^\prime,\theta^\prime)\}
    =\delta^I{}_J\delta(\sigma-\sigma^\prime)\delta(\theta-\theta^\prime), \nom\\
    &&\{\bx^I(\sigma,\theta),\vxi_J(\sigma^\prime,\theta)\}
    = - \{\vxi_J(\sigma,\theta), \bx^I(\sigma^\prime,\theta)\}= \delta^{I}{}_{J}\delta(\sigma-\sigma^\prime)\delta(\theta-\theta^\prime),
\label{2dsupercanonical}
\eeqa
where $\delta(\theta-\theta^\prime)$ is defined by $\theta+\theta^\prime$.

Next a Q-structure Hamiltonian functional $\vThe$
on $\Map(T[1]S^1, \calM)$ 
is constructed.
$\vThe$ is induced from the Q-structure 
Hamiltonian function $\Theta$ on $\calM$.
$\vThe$ is defined as follows:
\beq
    \vThe= \int_{T[1]S^1}  d\sigma d\theta\ \ev^* \Theta.
\eeq
Since the integration shifts degree by $1$,
$(\Map(T[1]S^1, \calM), \bbOmega, \vThe)$ is a QP manifold of degree $1$.
Therefore $\bbOmega$ and $\vThe$ define a Poisson structure.

Since $\calM =T^*[2](T^*[1]M)$ is a QP-manifold of degree $2$,
a Hamiltonian function $\Theta$ 
is of degree $3$ and
defines the Courant-Dorfman bracket from Ex.~\ref{Calgebroid}
by the derived bracket:
\beq
    [-,-]= \sbv{\sbv{-}{\Theta}}{-}.
\label{dbcourant}
\eeq
Since we do not consider any extra structure
except for a closed $3$-form $H$ in the original two dimensional system,
a Hamiltonian function 
is 
canonically taken as
\beq
    \ev^* \Theta=\veta^I \vxi_I   
          + \frac{1}{3!} H_{IJK}(\bx) \veta^I \veta^J \veta^K.
\eeq
$\sbv{\Theta}{\Theta}=0$ if and only if $H$ is a closed form.
Indeed the derived brackets of graded fields are obtained as follows:
\beqa
    &&\sbv{\sbv{\bx^I(\sigma,\theta)}{\vThe}}{\bx^J(\sigma^\prime,\theta^\prime)}=0,
\nonumber\\
  \noalign{\vskip 2mm}
    &&\sbv{\sbv{\bx^I(\sigma,\theta)}{\vThe}}{\bp_J(\sigma^\prime,\theta^\prime)}=\delta^I{}_J\delta(\sigma-\sigma^\prime)\delta(\theta-\theta^\prime),\nom\\
  \noalign{\vskip 2mm}
    && \sbv{\sbv{\bp_I(\sigma,\theta)}{\vThe}}{\bp_J(\sigma^\prime,\theta^\prime)}=-H_{IJK}(\bx(\sigma,\theta))\veta^K(\sigma,\theta)\delta(\sigma-\sigma^\prime)\delta(\theta-\theta^\prime).
\eeqa
The pullback of the embedding map $\hatj_*$ induce
the Poisson bracket on the original fields.
This corresponds to taking a projection to a submanifold
by making the projection of auxiliary fields,
\beqa
    &&
x^{(1)I}(\sigma) = 
p^{(0)}_I(\sigma)
= \eta^{(0)I} = 0,
\qquad \eta^{(1)I} = \del x^I,
\label{gaugefixing}
\eeqa
after the calculations of the graded Poisson bracket.
The Poisson brackets (\ref{2dpoisson}) 
of canonical quantities are
derived as
\beqa
    &&\{x^I(\sigma),x^J(\sigma^\prime)\}_{P.B.}= 
\hatj^* 
\sbv{\sbv{\bx^I(\sigma,\theta)}{\vThe}}{\bx^J(\sigma^\prime,\theta^\prime)},
\nonumber\\
  \noalign{\vskip 2mm}
    &&
\{x^I(\sigma),p_J(\sigma^\prime)\}_{P.B.}=
\hatj^* 
\sbv{\sbv{\bx^I(\sigma,\theta)}{\vThe}}{\bp_J(\sigma^\prime,\theta^\prime)},\nom\\
  \noalign{\vskip 2mm}
    && \{p_I(\sigma),p_J(\sigma^\prime)\}_{P.B.}
    = 
\hatj^* 
\sbv{\sbv{\bp_I(\sigma,\theta)}{\vThe}}{\bp_J(\sigma^\prime,\theta^\prime)}.
\eeqa
Generalized currents are converted to
super functions of degree of zero and degree of one
on $C^{\infty}(\calM)$ by 
$\hatj_*: J \longmapsto \bJ$, respectively:
\beq
    \bJ_{0(f)}(\sigma,\theta)=f(\bx),\quad {\rm and}\quad
    \bJ_{1(u,\alpha)}(\sigma,\theta) = \alpha_I(\bx) \veta^I + u^I(\bx) \bp_I.
\eeq
The original currents are recovered by the pullback of $\hatj$.
The derived brackets of these current functions are directly calculated 
from (\ref{2dsupercanonical}) as follows:
\beqa
    && \{\{\bJ_{0(f)}(\sigma,\theta),\vThe\},\bJ_{0(f^\prime)}(\sigma^\prime,\theta^\prime)\}
        =0,\nom\\
  \noalign{\vskip 2mm}
    && \{\{\bJ_{1(u,\alpha)}(\sigma,\theta),{\vThe}\},\bJ_{0(f^\prime)}(\sigma^\prime,\theta^\prime)\}
       =-u^{\prime I}\frac{\del f}{\del \bx^I}\delta(\sigma-\sigma^\prime)\delta(\theta-\theta^\prime),\nom\\
  \noalign{\vskip 2mm}
    &&\{\{\bJ_{1(u,\alpha)}(\sigma,\theta),{\vThe}\},\bJ_{1(u^\prime,\alpha^\prime)}(\sigma^\prime,\theta^\prime)\}
    =-\bJ_{1(\courant{(u,\alpha)}{(u^\prime,\alpha^\prime)})}(\sigma,\theta)
\delta(\sigma-\sigma^\prime)\delta(\theta-\theta^\prime),
\label{2dsupercommutation}
\eeqa
where
\beqa
    &&\bJ_{1(\courant{(u,\alpha)}{(u^\prime,\alpha^\prime)})}(\sigma,\theta)= 
\left[\left(u^J\frac{\del u^{\prime I}}{\del \bx^J}
    -u^{\prime J}\frac{\del u^I}{\del \bx^J}\right)\bp_I\right.\nom\\
    &&\hspace{4cm}\left.+\left(u^J\frac{\del \alpha^\prime_I}{\del \bx^J}
    -u^{\prime J}\frac{\del \alpha_I}{\del \bx^J}
    +u^{\prime J}\frac{\del \alpha_J}{\del \bx^I}
    +\alpha^\prime_J\frac{\del u^J}{\del \bx^I}
    +H_{JKI} u^{J} u^{\prime K}\right)\veta^I\right].\qquad\quad
\eeqa
Here Eq.~(\ref{dbcourant}) is used to derive the Courant bracket.
Eq.~(\ref{2dsupercommutation})
has the same relations as the current algebras 
(\ref{2dcurrentalgebra}) except for the anomaly terms.

The anomaly terms, which are proportional to the differentials of 
the delta function,
are obtained by the graded Poisson brackets of currents as
\beqa
    &&\{\bJ_{0(f)}(\sigma,\theta),\bJ_{0(f^\prime)}(\sigma^\prime,\theta^\prime)\}=0,\quad
    \{\bJ_{0(f)}(\sigma,\theta),\bJ_{1(u^\prime,\alpha^\prime)}(\sigma^\prime,\theta^\prime)\}=0,\nom\\
  \noalign{\vskip 2mm}
    &&\{\bJ_{1(u,\alpha)}(\sigma,\theta),\bJ_{1(u^\prime,\alpha^\prime)}(\sigma^\prime,\theta^\prime)\}
    =(\alpha_I u^{\prime I}+\alpha_{I}^{\prime} u^I)\delta(\sigma-\sigma^\prime)\delta(\theta-\theta^\prime).
\eeqa
The coefficients of the delta functions in the right hand sides are
the same as the coefficients of 
$\partial_{\sigma} \delta(\sigma-\sigma^\prime)$ in 
anomaly terms in the current algebra
(\ref{2dcurrentalgebra}).
Results are summarized to the following theorem:
\begin{theorem}
A current algebra {\rm (\ref{2dcurrentalgebra})} in two dimensions
has a realization in a QP manifold of degree $2$ on $T^*[2]T^*[1]M$
induced from the minimal symplectic realization 
$j: TM \oplus T^*M \longrightarrow T^*[2]T^*[1]M$.
The algebraic structure is calculated by the derived bracket 
of current super functions.
The anomaly cancellation condition is equivalent to the condition that 
currents super functions are commutative under the Q-structure.
\end{theorem}

\section{Generalized Current Algebras
in Three Dimensions}
\noindent
Current algebras in the previous sections
are generalized to three dimensions.
It is necessary to extend current algebras 
in three dimensions in order to contain currents
in a large class of field theories
such as the Chern-Simons theory with matters or
the Courant sigma model.

\subsection{Chern-Simons Theory with Matter}
\noindent
The action of complex scalar fields coupled with 
the Chern-Simons theory in three dimensions is 
\beqa
    S&=&\int_{X_3} d^3\sigma \Big[
    k_{IJ}(\partial_\mu x^I+f^I{}_K{}_L q_{\mu}^K x^L)
(\partial^{\mu} x^{\ast J}+f^J{}_M{}_N q^{\mu M} x^{\ast N})\nonumber\\
    &&\hspace{2cm}+V(k_{IJ}x^{\ast I} x^J)+\frac{k_{IJ}}{2}\varepsilon^{\mu\nu\rho}q^I_\mu\partial_\nu q^J_\rho
    +\frac{1}{3!}f_{IJK}\varepsilon^{\mu\nu\rho}q_{\mu}^Iq_{\nu}^Jq_{\rho}^K\Big].
\label{CSmatter}
\eeqa
where 
$X_3 = \Sigma \times \bR$ is a manifold in three dimensions,
$x^I$ is a complex scalar field and 
$q_{\mu}^K$ is a gauge field. 
$f_{IJK}$ is a structure constant of a Lie algebra, 
$k_{IJ}$ is a metric on a Lie algebra
Recently, this action is used to describe multiple M2-branes
\cite{Bagger:2006sk}
\cite{Bagger:2007jr}
\cite{Gustavsson:2007vu}
\cite{Aharony:2008ug}.
The action 
(\ref{CSmatter})
can be written in terms of differential forms as follows:
\beq
    S=\int_{X_3}k_{IJ}(dx^I+[q, x]^I)\wedge*(dx^{\ast J}+[q, x^\ast]^J)+*V(x)
    +\frac{k_{IJ}}{2}q^I\wedge dq^J+\frac{1}{3!}f_{IJK}q^I\wedge q^J\wedge q^K,
\eeq
where
\[
    [q, x]^I=f^I{}_{JK} q^J x^K,
\]
and $x^I(\sigma)$ and $q^I(\sigma) = d \sigma^{\mu} q^I_{\mu}(\sigma)$ 
are a $0$-form and a $1$-form, respectively. 
$*$ represents the Hodge star.
The canonical momenta are
\beqa
   &&\displaystyle p_{I}=\frac{\delta S}{\delta (\partial_0 x^I)}
    =k_{IJ}(\del_0 x^J+f^{J}{}_{KL}q_0^K x^L),\quad\nonumber\\
   &&\displaystyle p_{I}^\ast=\frac{\delta S}{\delta (\partial_0 x^{\ast I})}
    =k_{IJ}(\del_0 x^{\ast J}+f^{J}{}_{KL}q_0^K x^{\ast L}),\quad\nonumber\\
   &&\displaystyle\pi_{I}=\frac{\delta S}{\delta (\partial_0 q^I_{1})}=
    k_{IJ}q^J_{2}.
\eeqa
The symplectic structure is defined as 
\beq
    \omega=\int_{\Sigma}d^2\sigma \left(\delta x^I\wedge \delta p_{I}+
    \delta x^{\ast I}\wedge \delta p^\ast_{I}+k_{IJ}\delta q_1^I\wedge \delta q_{2}^J\right).
\label{symplecticCSM}
\eeq
The Hamiltonian of the system is following:
\beqa
    &&{\cal H}={\cal H}_0+q_0^I {J_1}_I,
\eeqa
where ${J_1}_I$ is a current of a gauge symmetry. 
Concrete expressions are
\beqa
    &&{\cal H}_0=k^{IJ}p^\ast_Ip_J
    -k_{IJ}\sum_{i=1}^2(\del_i x^I+f^I{}_{KL}q_i^K x^L)(\del_i x^{\ast J}+f^J{}_{MN}q_i^M x^{\ast N})-V(x),\\
    &&{J_1}_I=
k_{IJ}(\del_2q_1^J-\del_1q_2^J)-f_{IJK}q_1^Jq_2^K
- f{}_{IJ}{}^K(x^J p_K + x^{\ast J} p_K^\ast )
.
\eeqa
The Poisson bracket for gauge currents is as follows:
\beqa
\{{J_1}_I(\sigma),{J_1}_J(\sigma^\prime)\}_{P.B.}
=-f_{IJ}{}^K{J_1}_K(\sigma) \delta(\sigma-\sigma^\prime)\approx 0.
\eeqa
where we use the Jacobi identity:
\beqa
    k^{KN}(f_{KIL}f_{JMN}+f_{KLJ}f_{IMN}+f_{KIJ}f_{MLN})=0.
\eeqa
We can easily confirm that $J_1$ commutes with the Hamiltonian,
\beqa
    \{{J_1}_I(\sigma),{\cal H}(\sigma^\prime)\}_{P.B.}
=-q_0^J f_{IJ}^K
    {J_1}_K\delta(\sigma-\sigma^\prime)\approx 0.
\eeqa

\subsection{Courant Sigma Model}
\noindent
The Courant sigma model has been introduced in \cite{Ikeda:2002wh} 
and formulated by the AKSZ formalism in \cite{Roytenberg:2006qz},
which is a topological sigma model 
in the $1+2$-dimensional worldvolume.

This is a sigma model
on a three dimensional manifold $X_3$
to a vector bundle $E \longrightarrow M$.
This has the following action:
\begin{eqnarray}
    S=\int_{X_3} p_I \wedge d x^I +\frac{k_{AB}}{2}q^A \wedge dq^B
      -f_{1}{}^I{}_{A}(x) q^A\wedge p_I
+\frac{1}{3!} f_{2ABC}(x) q^A\wedge q^B\wedge q^C,
\label{3dcsm}
\end{eqnarray}
where $x^I$, $q^A$ and $p_I$ are 0-form, 1-form and 2-form 
on the base manifold $X_3$, respectively. 
$I, J, \cdots$ are indices of $M$, and 
$A, B, \cdots$ are indices of the fiber of $E$,
and $k_{AB}$ is a fiber metric on $E$.
By the construction, note that $k_{AB}$ is symmetric 
for $A,B$ and $f_{2ABC}(x)$ is skewsymmetric for $A,B,C$. 
More precisely, 
$x^I$ is a smooth map from $X_3$ to $M$, 
$q^A  \in \Gamma(T^*X_3 \otimes x^*(E))$ and
$p_I \in \Gamma(\wedge^2 T^*X_3 \otimes x^*(T^*M))$.

For consistency of the action, the structure functions
$f_{1}{}^I{}_{A}(x)$ and $f_{2ABC}(x)$ must have the following conditions 
(see \cite{Ikeda:2002wh}):
\beqa
    &&(1)\quad k^{AB}f_{1A}^If_{1B}^J=0,
\nom\\
    &&(2)\quad f_{1A}^J\frac{\partial f_{1B}^{I}}{\partial x^J}
    -f_{1B}^J\frac{\partial f_{1A}^{I}}{\partial x^J}
    +k^{CD}f_{1C}^If_{2DBA}=0,
\nom\\
    &&(3)\quad 
    f_{1D}^I\frac{\partial f_{2ABC}}{\partial x^I}
    -f_{1A}^I\frac{\partial f_{2BCD}}{\partial x^I}    
    +f_{1B}^I\frac{\partial f_{2CDA}}{\partial x^I}
    -f_{1C}^I\frac{\partial f_{2DAB}}{\partial x^I}
\nom\\
    &&\hspace{2cm}+k^{EF}(f_{2EAB}f_{2CDF}+f_{2EAC}f_{2DBF}
    +f_{2EAD}f_{2BCF})=0.
\label{CScon}
\eeqa
(\ref{CScon}) is equivalent to the condition that
a vector bundle $E$ is the Courant algebroid
\cite{Ikeda:2002wh}, \cite{Roytenberg:2006qz}.
If we take $f_{1}{}_{A}^I(x) = 0$ and $f_{2ABC}(x) = f_{2ABC}=$constant, 
the Courant algebroid reduces to a Lie algebra and the action 
(\ref{3dcsm}) reduces to the Chern-Simons gauge theory plus a BF theory.
This model contains a lot of known models, such as 
the Chern-Simons gauge theory and 
the Rozansky-Witten theory \cite{Qiu:2009zv}.

We set $X_3 = \Sigma_2 \times \bR$, 
and
$q^A = q^A_{\mu} d \sigma^{\mu}$
and $p_I = \frac{1}{2}p_{I \mu\nu}d \sigma^{\mu} \wedge d \sigma^{\nu}$.
Since the canonical momenta yield
\beq
    \displaystyle\pi_{I}=\frac{\delta S}{\delta (\partial_0 x^I)}=p_{I12}=-p_{I21},\quad
    \displaystyle\pi_{A}=\frac{\delta S}{\delta (\partial_0 q^A_{1})}=
    k_{BA}q^B_{2},
\eeq
the symplectic structure is written as 
\beq
    \omega=\int_{\Sigma_2}d^2\sigma \left(\delta x^I\wedge \delta p_{I12}+k_{AB}\delta q_1^A\wedge \delta q_{2}^B\right).
\label{symplecticCSM}
\eeq
The Hamiltonian of the system is following:
\beqa
{\cal H}&=&
\int_{\Sigma_2} d^2\sigma 
   \left(\pi_{I}\partial_0 x^I+\pi_{A}\partial_0 q^{A}_{1}-{\cal L}\right)\nom\\
 &=&\int_{\Sigma_2} d^2\sigma
   {\Big (}-p_{I01}(\partial_2 x^I-f^{I}_{1A}q^A_2)+p_{I02}(\partial_1 x^I-f^{I}_{1A}q^A_1)\nom\\
 &&\hspace{3cm}-q^A_0\left(k_{AB}(\partial_1 q^B_{2}-\partial_2 q^B_{1})-f_{1A}^{I}p_{I12}+f_{2ABC}q^B_1q^C_2 \right){\Big )}.
\eeqa
Since the fields $p_{I01},p_{I02},q^A_0$ are not dynamical variables,
the following quantities are referred as currents, 
\begin{eqnarray}
   J_{1i}^I(\sigma)&=&\partial_{i} x^I(\sigma)-f_{1A}^I(x(\sigma)) q_{i}^A(\sigma),\nom\\
   J_{2A}(\sigma)&=&k_{AB}(\partial_1q^B_2({\sigma})-\partial_2q^B_1({\sigma}))
   -f_{1A}^I(x({\sigma}))p_{I12}({\sigma})
   +f_{2ABC}(x({\sigma}))q_1^B({\sigma})q_2^C({\sigma}),\quad
\label{currentCSM}
\end{eqnarray}
where $i = 1, 2$.
In fact, they are constraints
$J_{1i}^I(\sigma)\approx 0$ and $J_{2A}(\sigma)\approx 0$.
The Hamiltonian
is expressed in term of these currents as follows:
\beq
    {\cal H}=\int_{\Sigma_2} d^2\sigma
(-p_{I01}J^{I}_{12}+p_{I02}J_{11}^{I}-q_0^AJ_{2A}).
\eeq

For consistency of the system, 
the Poisson brackets of constraints must be closed on the constraint subspace.
The current algebra takes the following commutation relations:
\beqa
  &&\{J_{1i}^I({\sigma}),J_{1j}^J({\sigma}^\prime)\}_{P.B.}=0, 
\nonumber \\
  &&\{J_{2A}({\sigma}),J_{1i}^I({\sigma^\prime})\}_{P.B.}
     =\frac{\partial f_{1A}^I}{\partial x^J} (x(\sigma)) J_{1i}^J ({\sigma})
\delta^2({\sigma}-{\sigma}^\prime),\nom\\
  \noalign{\vskip 2mm}
   &&\{J_{2A}({\sigma}),J_{2B}({\sigma}^\prime)\}_{P.B.}
     ={\Big (}\frac{\partial f_{2ABC}}{\partial x^I}(x(\sigma))
(J_{11}^I ({\sigma})q_2^{C}({\sigma})-J_{12}^I ({\sigma})q_1^{C}({\sigma}))\nom\\
   &&\hspace{7cm}
+f_{2ABC}(x({\sigma}))k^{CD}J_{2D}({\sigma}){\Big )}\delta^2({\sigma}-{\sigma}^\prime),
\eeqa
with Eq.~(\ref{CScon}). 


\subsection{Generalized Current Algebras in Three Dimensions}
\noindent
We generalize a current algebra in three dimensional worldvolume 
as a generalization of Alekseev and Strobl, and
in order to contain the currents
of the matters coupled with the Chern-Simons theory or
the Courant sigma model in the previous subsections.

Let us take the worldvolume $X_3 = \Sigma_2 \times \bR$ and 
a target vector bundle $E$.
Since the Courant sigma model is generalized by introducing a closed
$4$-form \cite{Hansen:2009zd},
the symplectic structure (\ref{symplecticCSM}) 
can be generalized by introducing a closed $4$-form
$H = \frac{1}{4!} H_{IJKL}(x) d x^I d x^J d x^K d x^L$ 
on a target space $M$, which is
\beq
    \omega=\int_{\Sigma_2} d^2\sigma 
    (\delta x^I\wedge \delta p_{I12}+k^{AB}\delta q_1^A\wedge \delta q_2^B)
    +\frac{1}{4}\int_{\Sigma_2} d^2\sigma H_{IJKL}\epsilon^{ij}
    \del_i x^I\del_j x^J \delta x^K\wedge \delta x^L,
\eeq
where $i,j=1,2$ and $\epsilon^{12}=1,\epsilon^{21}=-1$ (and others are $0$).
Then the Poisson brackets of canonical quantities are 
\beqa
    &&\{x^I(\sigma),p_{Jij}(\sigma^\prime)\}_{P.B.}=\epsilon_{ij}\delta^I_{J}\delta^2(\sigma-\sigma^\prime),\quad
    \{q^A_i(\sigma),q^B_j(\sigma^\prime)\}_{P.B.}=\epsilon_{ij}k^{AB}\delta^2(\sigma-\sigma^\prime),\nom\\
    &&\{p_{Iij}(\sigma),p_{Jkl}(\sigma^\prime)\}_{P.B.}=-\frac{1}{2}\epsilon_{ij}\epsilon_{kl}H_{IJKL}(x)
    \epsilon^{mn}\del_{m}x^K\del_{n}x^L\delta^2(\sigma-\sigma^\prime).
\eeqa
The mass dimensions of each quantities are
chosen so that $\omega$ is dimensionless \cite{Bonelli:2005ti}.
Since $\dim[\sigma]=-1$ and $\dim[\partial]=1$,
the mass dimensions of the canonical conjugates are taken as
$\dim[x^I]=0$,
$\dim[q^A]=1$ 
and
$\dim[p_I]=2$.

If it is assumed that each current has the homogeneous mass dimension, 
the following 
three currents are most general forms of mass dimension zero, one and two,
respectively:
\beqa
    &&J_{0(f)}({\sigma})=f(x(\sigma)), 
\nonumber \\
&&    J_{1i(\alpha,u)}({\sigma})=
    \alpha_{I}(x({\sigma}))\partial_i x^I({\sigma})+u_{A} (x({\sigma}))q_i^A({\sigma}),\nom\\
  \noalign{\vskip 2mm}
    &&J_{2ij(G,K,F,B,E)}({\sigma})
    =\epsilon_{ij}\epsilon^{kl}
    \left(\frac{1}{2}G^I(x({\sigma}))p_{Ikl}({\sigma})+K_A(x({\sigma}))\partial_k q_l^A({\sigma})
    +\frac{1}{2}F_{AB}(x({\sigma}))q_k^A({\sigma})q_l^B({\sigma})\right.\nom\\
    &&\hspace{4.5cm}\left.+\frac{1}{2}B_{IJ}(x({\sigma}))\partial_k x^I({\sigma})\partial_l x^J({\sigma})
      +E_{AI}(x({\sigma}))\partial_k x^I({\sigma}) q_l^A({\sigma})\right).
\eeqa

If the canonical quantities are rewritten 
by differential forms $q^A=q^A_i d\sigma^i$,\ 
$p_I=\frac{1}{2}p_{Iij}d\sigma^i\wedge d\sigma^j$
on the space direction $\Sigma$,
the canonical commutation relations are
\beqa
&&    \{x^I(\sigma),p_{J}(\sigma^\prime)\}=\delta^I_J\delta^2(\sigma-\sigma^\prime),\quad
    \{q^A(\sigma),q^B(\sigma^\prime)\}=-k^{AB}\delta^2(\sigma-\sigma^\prime),
\nonumber \\
&&    \{p_I(\sigma),p_J(\sigma^\prime)\}
    =-\frac{1}{2}H_{IJKL}d x^K\wedge dx^L \delta^2(\sigma-\sigma^\prime)
\label{3dcanonical}
\eeqa
Generalized currents in differential forms are written by
\beq
    J_{0(f)}(\sigma)=f,\quad J_{1(\alpha,u)}(\sigma)=\alpha+u,
    \quad J_{2(G,K,F,B,E)}(\sigma)=G+K+F+B+E,
\label{3dcurrent}
\eeq
respectively. Here $\alpha,u,G,K,F,H,E$ are defined as forms:
\beqa
   &&\alpha=\alpha_I(x(\sigma)) dx^I,\quad u=u_A(x(\sigma)) q^A,\quad
   G=G^I(x(\sigma))p_I(\sigma),\quad K=K_A(x(\sigma)) dq^A, \nom\\
   &&F=\frac{1}{2}F_{AB}(x(\sigma))q^A(\sigma)\wedge q^B(\sigma),\quad
     B=\frac{1}{2}B_{IJ}(x(\sigma))dx^I(\sigma)\wedge dx^J(\sigma),\nom\\
   &&E=E_{AI}(x(\sigma))dx^I(\sigma)\wedge q^A(\sigma).
\eeqa
The Poisson brackets of currents 
are directly calculated by using (\ref{3dcanonical}) as
the following forms:
\beqa
        &&\sbv{J_{0(f)}(\sigma)}{J_{0(f^\prime)}(\sigma^\prime)}=0,\nom
\label{3dcurrentalgebra}\\
        &&\sbv{J_{1(\alpha,u)}(\sigma)}{J_{0(f^\prime)}(\sigma^\prime)}=0,\nom\\
        &&\sbv{J_{2(G,K,F,B,E)}(\sigma)}{J_{0(f^\prime)}(\sigma^\prime)}
          =-i_{G}df^\prime\delta^2(\sigma-\sigma^\prime),\nom\\
        &&\sbv{J_{1(\alpha,u)}(\sigma)}{J_{1(\alpha^\prime,u^\prime)}(\sigma^\prime)}
          =-\langle u,u^\prime \rangle\delta^2(\sigma-\sigma^\prime),\nom\\
        &&\sbv{J_{2(G,K,F,B,E)}(\sigma)}{J_{1(\alpha^\prime,u^\prime)}(\sigma^\prime)}\nom\\
        &&\hspace{2cm}=-J_{1(\bar\alpha,\bar u)}\delta^2(\sigma-\sigma^\prime)+({\it i}_G\alpha^\prime-\langle u^\prime,K\rangle)d\delta^2(\sigma-\sigma^\prime),\nom\\
        &&\sbv{J_{2(G,K,F,B,E)}(\sigma)}{J_{2(G^\prime,K^\prime,F^\prime,B^\prime,E^\prime)}(\sigma^\prime)}\nom\\
        &&\hspace{5mm}=-J_{2(\bar G,\bar K,\bar F,\bar B,\bar E)}\delta^2(\sigma-\sigma^\prime)\nom\\
        &&\hspace{5mm}-({\it i}_G (E^\prime+B^\prime)+{\it i}_G^\prime (E+B)
          +\langle E^\prime+F^\prime,K\rangle+\langle E+F,K^\prime\rangle)\wedge d\delta^2(\sigma-\sigma^\prime),
\label{3dcurrentalgebra}
\eeqa
and
\beqa
    &&\bar \alpha=({\it i}_{G} d+d{\it i}_{G})\alpha^\prime+\langle E-dK,u^\prime\rangle,\quad
    \bar u={\it i}_{G} du^\prime +\langle F, u^\prime\rangle, \nom
\\
    &&\bar G=[G,G^\prime], \nom\\
    &&\bar K={\it i}_{G}d K^\prime-{\it i}_{G^\prime}d K+{\it i}_{G^\prime} E+\langle F, K^\prime\rangle, \nom\\
    &&\bar F={\it i}_{G}d F^\prime-{\it i}_{G^\prime}d F+\langle F, F^\prime\rangle, \nom\\ 
    &&\bar B=(d{\it i}_{G}+{\it i}_{G}d)B^\prime-{\it i}_{G^\prime}d B+\langle E, E^\prime\rangle
    +\langle K^\prime, dE\rangle-\langle dK,E^\prime\rangle+{\it i}_{G^\prime}{\it i}_{G} H, \nom\\
    &&\bar E=(d{\it i}_{G}+{\it i}_{G}d)E^\prime-{\it i}_{G^\prime}d E+\langle E, F^\prime\rangle
    -\langle E^\prime, F\rangle+\langle dF, K^\prime\rangle-\langle dK, F^\prime\label{3bE}
\rangle, 
\label{3dalgebra}
\eeqa
where all the terms are evaluated by $\sigma^{\prime}$.
Here $[ -,- ]$ is a Lie bracket on $TM$, $i_G$ is an interior product
with respect to a vector field $G$ and
$\langle -,- \rangle$ is the graded bilinear form on the fiber of $E$
with respect to the metric $k^{AB}$.
For example,
$\langle q^A,q^B \rangle=k^{AB}, 
\langle q^A\wedge q^B,q^C \rangle=q^A k^{BC}-q^B k^{AC}$, etc.
Component expressions of 
the equations (\ref{3dcurrentalgebra}) and (\ref{3dalgebra}) appears
in the Appendix.

\subsection{Current Algebras and QP structures of degree $3$}
\noindent
We point out that 
Current algebras in the previous subsection are constructed
from a QP structure of degree $3$ on $\calM = T^*[3]T^*[2]E[1]$
in this subsection.

Let us extend the space direction of the worldvolume to 
the supermanifold $T[1]\Sigma_2$
with a local coordinate $(\sigma^i, \theta^i)$.
Let $(T^*[3]T^*[2]E[1], \Omega, \Theta)$ be a QP manifold of degree $3$.
A target space $T^*E$ has 
a natural symplectic realization
$j: T^*E \longrightarrow T^*[3]T^*[2]E[1]$,
where $j: (x^I, q^A, \frac{\partial}{\partial x^I}, 
0, d q^A, d x^I)
 \longmapsto (x^I, q^A, p_I, \xi_I, \eta^A, \chi^I)$.
This map and a natural pairing of $TM$ and $T^*M$
induce an embedding map of mapping space of superfields 
$\hatj: (x^I, q^A, p_I, 
0, d q^A, d x^I)
 \longmapsto (\bx^I, \bq^A, \bp_I, \vxi_I, \veta^A, \vchi^I)$.
Let us consider the space of smooth map from 
$T[1]\Sigma_2$ to $T^*[3](T^*[2]E[1])$,
which is denoted by 
$\Map(T[1]\Sigma_2, T^*[3](T^*[2]E[1]))$.

Let us consider a local coordinate expression.
Let $\bx^I(\sigma,\theta) = x^I(\sigma) + \theta^i x_i^{(1)I}(\sigma)
+ \frac{1}{2} \theta^i \theta^j x_{ij}^{(2)I}(\sigma)
$ be a smooth map from $T[1]\Sigma_2$ to $M$.
$\bq^A(\sigma,\theta) \in \Gamma(T^*[1]\Sigma_2
\otimes \bx^*(E[1]))$ is a superfield of degree $1$,
$\bq^A(\sigma,\theta) = q^{(0)A}(\sigma) + \theta^i q_i^{A}(\sigma)
+ \frac{1}{2} \theta^i \theta^j q_{ij}^{(2)A}(\sigma)$,
which contains $q^A$ as a component.
A superfield containing $p_I$ is 
$\bp_I(\sigma,\theta) \in \Gamma(T^*[1]\Sigma_2
\otimes \bx^*(T^*[2]M))$,
$\bp_I(\sigma,\theta) = p^{(0)}_I(\sigma) + \theta^i p_{Ii}^{(1)}(\sigma)
+ \frac{1}{2} \theta^i \theta^j p_{Iij}(\sigma)$.
%
A graded symplectic form 
$\bbOmega$ of degree $1$ is defined from $\Omega$ as
\beq
    \bbOmega
=\int_{T[1] \Sigma_2} 
d^2\sigma d^2 \theta \ \ev^* \Omega
=\int_{T[1] \Sigma_2} 
d^2\sigma d^2 \theta
(\delta \bx^I\wedge\delta\vxi_I
    +k_{AB}\delta\bq^A\wedge\delta \veta^{B}
+\delta \bp_I\wedge\delta \vchi^I),
\eeq
where 
$\vxi_I(\sigma,\theta) \in \Gamma(T^*[1]\Sigma_2 \otimes \bx^*(T^*[3]M))$,
$\veta^A(\sigma,\theta) 
\in \Gamma(T^*[1]\Sigma_2\otimes \bx^*(T^*[3]E[1]))$, 
 and
$\vchi^I(\sigma,\theta) \in \Gamma(T^*[1]\Sigma_2 \otimes \bx^*(T^*[3]T^*[2]M))$ 
are 'canonical conjugates' of $\bx^I$, $\bq^A$ and $\bp_I$
with respect to $\bbOmega$.
The graded Poisson brackets on 'canonical conjugates' 
are
\beqa
    &&\{\bx^I(\sigma,\theta),\vxi_J(\sigma^\prime,\theta^\prime)\}=\delta^I{}_J\delta^2(\sigma-\sigma^\prime)\delta^2(\theta-\theta^\prime),\nom\\    
    &&\{\bq^A(\sigma,\theta),\veta^B(\sigma^\prime,\theta^\prime) \}= k^{AB}\delta^2(\sigma-\sigma^\prime)\delta^2(\theta-\theta^\prime),\nom\\
    &&\{\bp_I(\sigma,\theta),\vchi^J(\sigma^\prime,\theta^\prime)\}= \delta^J{}_I\delta^2(\sigma-\sigma^\prime)\delta^2(\theta-\theta^\prime),\quad
\eeqa
where $\delta^2(\theta-\theta^\prime)
=(\theta_1+\theta_1^\prime)(\theta_2+\theta_2^\prime)$. 

A Q-structure Hamiltonian functional $\vThe$
is induced from the Q-structure 
Hamiltonian function $\Theta$ on $T^*[3]T^*[2]E[1]$.
$\vThe$ is defined as follows:
\beq
    \vThe= \int_{T[1]\Sigma}  d^2\sigma d^2\theta\ \ev^* \Theta.
\eeq
Since the integration shifts degree by $2$,
$(\Map(T[1]S^1, \calM), \bbOmega, \vThe)$ is a QP manifold of degree $1$
and $\bbOmega, \vThe$ define a Poisson structure.

A Q-structure Hamiltonian function $\Theta$ is of degree $4$.
Since no background structure except for a closed $4$-form $H$
is considered, the local coordinate expression of
the pullback of $\Theta$ is canonically taken as
\beq
    \ev^* \Theta=  \vchi^I\vxi_I  
          +\frac{1}{2}k_{AB}\veta^A\veta^B+ 
\frac{1}{4!}
H_{IJKL}(\bx) \vchi^I \vchi^J \vchi^K \vchi^L.
\eeq
The derived brackets for fundamental superfields are
\beqa
    &&\sbv{\sbv{\bx^I(\sigma,\theta)}{\Theta}}{\bp_J(\sigma^\prime,\theta^\prime)}
    =-\sbv{\sbv{\bp_I(\sigma,,\theta)}{\Theta}}{\bx^J(\sigma^\prime,\theta^\prime)}
    =\delta^I{}_J\delta^2(\sigma-\sigma^\prime)\delta^2(\theta-\theta^\prime),\nom\\
    &&\sbv{\sbv{\bq^A(\sigma,\theta)}{\Theta}}{\bq^B(\sigma^\prime,\theta^\prime)}
    =- k^{AB}\delta^2(\sigma-\sigma^\prime)\delta^2(\theta-\theta^\prime),\quad\nom\\
    &&
\sbv{\sbv{\bp_I(\sigma,\theta)}{\Theta}}{\bp_J(\sigma^\prime,\theta^\prime)}
    =-\frac{1}{2}H_{IJKL}\vchi^K\vchi^L\delta^2(\sigma-\sigma^\prime) \delta^2(\theta-\theta^\prime),
\eeqa
These derive the Poisson brackets for canonical conjugates by 
the pullback of the embedding map $\hatj$:
\beqa
    &&
\sbv{x^I(\sigma,\theta)}{p_J(\sigma^\prime,\theta^\prime)}_{P.B.}=
\hatj^*
\sbv{\sbv{\bx^I(\sigma,\theta)}{\Theta}}{\bp_J(\sigma^\prime,\theta^\prime)},\nom\\
    &&
\sbv{q^A(\sigma,\theta)}{q^B(\sigma^\prime,\theta^\prime)}_{P.B.}=
\hatj^*
\sbv{\sbv{\bq^A(\sigma,\theta)}{\Theta}}{\bq^B(\sigma^\prime,\theta^\prime)}
,\quad\nom\\
    &&
\sbv{p_I(\sigma)}{p_J(\sigma^\prime)}_{P.B.}=
\hatj^*
\sbv{\sbv{\bp_I(\sigma,\theta)}{\Theta}}{\bp_J(\sigma^\prime,\theta^\prime)},
\eeqa
This is realized by 
the 'gauge fixings' of auxiliary fields,
which are carried out after the calculations as
\beqa
    &&
x_i^{(1)I} = x_{ij}^{(2)I} = 
q^{(0)A} = 
q_{ij}^{(2)A} = 
p^{(0)}_I = p_{Ii}^{(1)} =
\chi^{(0)I} 
= 0,
\nom\\
&& 
\chi_i^{(1)I} = \partial_i x^I,\nom\\
&& 
\eta_{ij}^{(2)A} = \partial_i q_j^A - \partial_j q_i^A.
\label{gaugefix3d}
\eeqa
Since functions on the supermanifold
$C^{\infty} (T[1]\Sigma_2)$ can be identified to 
the exterior algebra ${\bigwedge}^\bullet T^* \Sigma_2$,
the original Poisson bracket (\ref{3dcanonical}) is obtained.

$\hatj$ maps generalized currents (\ref{3dcurrent}) 
to current super functions of degree $0, 1$ and $2$ as
\beqa
    && \hatj \ J_{0(f)} =
\bJ_{0(f)}(\sigma,\theta)=f(\bx(\sigma,\theta)),\nonumber \\
    && 
\hatj \ J_{1(\alpha,u)} =
\bJ_{1(\alpha,u)}({\sigma,\theta})=
\alpha_{I}(\bx({\sigma,\theta}))\vchi^I(\sigma,\theta)+u_{A} (\bx({\sigma,\theta}))\bq^A({\sigma,\theta})
,\nom\\
  \noalign{\vskip 2mm}
    &&
\hatj \ J_{2(G,K,F,H,E)} =
\bJ_{2(G,K,F,H,E)}({\sigma,\theta})\nom\\
    &&\hspace{1cm}=
    \left(G^I(\bx({\sigma,\theta}))\bp_{I}({\sigma,\theta})+K_A(\bx({\sigma,\theta}))\veta^A({\sigma,\theta})
    +\frac{1}{2}F_{AB}(\bx({\sigma,\theta}))\bq^A({\sigma,\theta})\bq^B({\sigma,\theta})\right.\nom\\
    &&\hspace{2.5cm}\left.+\frac{1}{2}B_{IJ}(\bx({\sigma,\theta}))\vchi^I({\sigma,\theta})\vchi^J({\sigma,\theta})
      +E_{AI}(\bx({\sigma,\theta}))\vchi^I({\sigma,\theta}) \bq^A({\sigma,\theta})\right).
\eeqa
Straightforward calculations show that the derived brackets describe the 
correct commutation relations (\ref{3dcurrentalgebra})
of current algebras:
\beqa
    && \sbv{\sbv{\bJ_{0(f)}}{\vThe}}{\bJ_{0(f^\prime)}}=0,\nom\\
    && \sbv{\sbv{\bJ_{1(u,\alpha)}}{\vThe}}{\bJ_{0(f^\prime)}}=0,\nom\\
    && \sbv{\sbv{\bJ_{2(G,K,F,H,E)}}{\vThe}}{\bJ_{0(f^\prime)}}
       =-G^{I}\frac{\del f^\prime}{\del \bx^I}\delta^2(\sigma-\sigma^\prime)\delta^2(\theta-\theta^\prime),\nom\\
    && \sbv{\sbv{\bJ_{1(u,\alpha)}}{\vThe}}{\bJ_{1(u^\prime,\alpha^\prime)}}
    =-k^{AB}u_A u_B^\prime\delta^2(\sigma-\sigma^\prime)\delta^2(\theta-\theta^\prime),\nom\\
    &&\sbv{\sbv{\bJ_{2(G,K,F,B,E)}}{\vThe}}{\bJ_{1(u^\prime,\alpha^\prime)}}
    =-\bJ_{1(\bar u,\bar \alpha)}\delta^2(\sigma-\sigma^\prime)\delta^2(\theta-\theta^\prime),\nom\\
    &&\sbv{\sbv{\bJ_{2(G,K,F,B,E)}}{\vThe}}{\bJ_{2(G^\prime,K^\prime,F^\prime,B^\prime,E^\prime)}}
    =-\bJ_{2(\bar G, \bar K, \bar F,\bar B,\bar E)}\delta^2(\sigma-\sigma^\prime)\delta^2(\theta-\theta^\prime),
\eeqa
where $\bar \alpha,\bar u, \bar G, \bar K, \bar F, \bar H, \bar E$ are 
in Eq. (\ref{3bE}).

The commutators of current functions with 
respect to the P-structure are 
\beqa
    &&\{\bJ_{0(f)},\bJ_{0(f^\prime)}\}=0,\quad
      \{\bJ_{1(u,\alpha)},\bJ_{0(f^\prime)}\}=0,\quad
      \{\bJ_{2(G,K,F,B,E)},\bJ_{0(f^\prime)}\}=0,
\nonumber \\
    &&\{\bJ_{1(u,\alpha)},\bJ_{1(u^\prime,\alpha^\prime)}\}=0,
 \nonumber \\
&&      \{\bJ_{2(G,K,F,B,E)},\bJ_{1(u^\prime,\alpha^\prime)}\}
      =(G^{I}\alpha^\prime_I-k^{AB}K_A u_B^\prime)\delta^2(\sigma-\sigma^\prime)\delta^2(\theta-\theta^\prime),\quad\nom\\
    && \{\bJ_{2(G,K,F,B,E)},\bJ_{2(G^\prime,K^\prime,F^\prime,B^\prime,E^\prime)}\}
      =
\left[(G^JB^\prime_{JI}+G^{\prime J}B_{JI}
             +k^{AB}(K_A E_{BI}^\prime+E_{AI}K_{B}^\prime))\vchi^I\right.\nom\\
    &&
\hspace{2cm}   
\left.+(G^IE^\prime_{AI}+G^{\prime I}E_{AI}
                +k^{BC}(K_B F_{AC}^\prime+F_{AC}K_{B}^\prime))\bq^A \right]
       \delta^2(\sigma-\sigma^\prime)\delta^2(\theta-\theta^\prime),\qquad\quad
\eeqa
which derive the correct coefficients of anomaly terms 
in (\ref{3dcurrentalgebra}).

We have obtained the following result.
\begin{theorem}
A current algebra {\rm (\ref{3dcurrentalgebra})} in three dimensions
has a realization as a QP manifold of degree $3$
on $T^*[3](T^*[2]E[1])$, i.e. a
Lie algebroid up to homotopy on the vector bundle $T^*E$.
The anomaly cancellation condition is equivalent to the condition 
that
currents super functions are commutative under the Q-structure.
\end{theorem}

\section{QP Structures of Current Algebras in $n$ Dimensions}
\noindent
We have seen that there are correspondences between 
generalized current algebras and QP manifolds in two and 
three dimensions.
Generalizations to $n$ dimensions is straightforward.
A generalized current algebra in $n$ dimensional worldvolume has 
a structure of a QP manifold of degree $n$. 

\subsection{Current Algebras in $n$ Dimensions}
\noindent
Let $X_n = \Sigma_{n-1} \times \bR$ be a manifold in $n$ dimensions
and a target space is a symplectic manifold 
in $d$ dimensions.
$\bR$ is the time direction and $\Sigma_{n-1}$ is the $n-1$ dimensional space.
Let $q^{A^{(i)}}_i(\sigma)$ be a canonical quantity,
which is a $i$-form, where $i = 0, 1, \cdots, n-1$.
The space of $q^{A^{(i)}}_i(\sigma)$ is $\Map(\Sigma_{n-1}, T^*E)$, 
where 
$E = \bigoplus_{i=1}^{\floor{\frac{n}{2}}} E_i$ is a vector bundle on $M$,
where $x^{A^{(0)}} = q^{A^{(0)}}_0$ is a map from $\Sigma_{n-1}$ to $M$
and
$q^{A^{(i)}}_i$ is a section of 
$\wedge^i T^* \Sigma_{n-1} \oplus x^*(E_i)$ 
for $1 \leq i \leq \floor{\frac{n}{2}}$.
The mass dimensions are $\dim[\sigma]=-1$ and $\dim[\partial]=1$.
$q^{A^{(i)}}_i(\sigma)$ has $\dim[q^{A^{(i)}}_i]=i$
in order to make the symplectic form $\omega$ dimensionless.

Let us consider the Poisson brackets for canonical conjugates:
\beqa
    &&\sbv{q^{A^{(i)}_1}_i(\sigma)}{q^{A^{(j)}_2}_{j}(\sigma^\prime)}_{P.B.}
    =(-1)^{ij+1}\sbv{q^{A^{(j)}_2}_j(\sigma)}{q^{A^{(i)}_1}_{i}(\sigma^\prime)}_{P.B.} \nonumber \\
&& \hspace{3cm}= (-1)^ik_i^{A^{(i)}_1A^{(j)}_2}\delta_{i, n-j-1}\delta^{n-1}(\sigma-\sigma^\prime)\quad {\rm for}\quad 0 \leq i \leq j < n-1, \\
    &&\sbv{q^{A^{(0)}_1}_{n-1}(\sigma)}{q^{A^{(0)}_2}_{n-1}(\sigma^\prime)}_{P.B.}
\nonumber\\
    && \qquad =-\frac{1}{(n-1)!}H^{A^{(0)}_1A^{(0)}_2}{}_{A^{(0)}_3\cdots A^{(0)}_{n+1}}dq^{A^{(0)}_3}_{0}\wedge\cdots\wedge dq^{A^{(0)}_{n+1}}_{0}
    \delta^{n-1}(\sigma-\sigma^\prime), 
\eeqa
and the other commutation relations are $0$,
where 
$H$ is a closed $n+1$-form and
$k_i^{A^{(i)}_1A^{(i)}_2}=(-1)^{(i+1)(n-i)}k_i^{A^{(j)}_2A^{(i)}_1}$ 
is a metric on the subspace of $q^{A^{(i)}_1}_i$ and $q^{A^{(j)}_2}_{j}$
and $A^{(i)}=A^{(n-i-1)}$.
Note that if $n$ is odd, the $i=j$ term
\beqa
    &&\sbv{q^{A^{(m)}_1}_m(\sigma)}{q^{A^{(m)}_2}_{m}(\sigma^\prime)}
    = (-1)^mk_m^{A^{(m)}_1A^{(m)}_2}\delta^{n-1}(\sigma-\sigma^\prime),
\eeqa
is nonzero, where $m= \frac{n-1}{2}$.

Currents 
$J_l(q^{A^{(i)}}_i(\sigma), d_{\sigma} q^{A^{(i)}}_{i}(\sigma))$ 
of mass dimensions $l$ are considered,
where $l = 0, 1, \cdots, n-1$.
The Poisson bracket of currents 
$\sbv{J_l(q^{A^{(i)}}_i(\sigma), d_{\sigma} q^{A^{(i)}}_{i}(\sigma))}
{J_{l^\prime}(q^{A^{(i)}}_i(\sigma^{\prime}), d_{\sigma^{\prime}} q^{A^{(i)}}_{i}(\sigma^{\prime}))}_{P.B.}$ has two terms, a commutator term 
and an anomaly term.

\subsection{QP-structures}

The space of worldvolume $\Sigma_{n-1}$ is extended to 
a supermanifold $T[1]\Sigma_{n-1}$ with a local coordinate 
$(\sigma^i, \theta^i)$.
A canonical quantity $q^{A^{(i)}}_i$ is extended to a superfield
$\bq^{A^{(i)}}_i 
$.
A superfield $\bq^{A^{(i)}}_i$
of degree $i$ ($i=0,\cdots, n-1$) contains
an original field $q^{A^{(i)}}_i$ as the $i$-th part.
The target vector bundle is extended to 
the double graded vector bundle
$\calM = T^*[n] \left(T^*[n-1] \left( 
\bigoplus_{i=1}^{\floor{\frac{n}{2}}} E_i[i] \right) \right)$,
which is a QP-manifold of degree $n$
with a QP structure $(\Omega, \Theta)$.
Note that the original canonical commutation relations of 
the original canonical quantities $q^{A^{(i)}}_i$ are not assumed
in this stage.

The original target space $T^*E$ is canonically embedded 
by the graded symplectic realization $j: T^*E \longrightarrow \calM$.
This induces a map $\hatj$ on a mapping space
$\hatj:(q^{A^{(i)}_1}_i, 
0, d_{\sigma} q^{A^{(i)}_1}_i)
\longmapsto
(\bq^{A^{(i)}_1}_i, 
\veta^{A^{(n-1)}_1}_{n}, \veta^{A^{(i)}_1}_{i+1})$,
where $\veta^{A^{(i)}_1}_{i+1} \equiv (-1)^{i(n-i)}k_i^{A^{(i)}_1A^{(i)}_2}
\veta_{A^{(i)}_2\, i+1}$.
Then the total space of a 'super' current algebras is
$C^{\infty}(\Map(T[1]\Sigma_{n-1}, \calM))$.
Superfields $\veta_{A^{(i)}\, i+1}$ of degree $i+1$
derived from fiber local coordinates 
are introduced ($i=0,\cdots, n-1$) 
as 'canonical conjugates' and
the odd symplectic structure (P-structure) is defined 
from graded symplectic structure $\Omega$ of degree $n$ 
on $\calM$
as
\beq
    \bbOmega =
\int_{T[1]\Sigma_{n-1}} 
d^{n-1} \sigma d^{n-1} \theta \ \Phi^* \Omega
= 
\int_{T[1]\Sigma_{n-1}} 
d^{n-1} \sigma d^{n-1} \theta \ 
\left(
\sum_{i=0}^{n-1} 
    \delta\bq^{A^{(i)}}_i
\wedge \delta \veta_{A^{(n-i-1)}\, {n-i}}
\right)
,
\eeq
where $A^{(i)}=A^{(n-i-1)}$.
Thus the graded Poisson brackets for 'canonical conjugates' are
\beqa
    &&\sbv{\bq^{A_1^{(i)}}_i(\sigma,\theta)}{\veta_{A_2^{(j)}\, {j+1}}(\sigma^\prime,\theta^\prime)}
    =- (-1)^{i(j+1)} 
\sbv{\veta_{A_2^{(j)}\, {j+1}}(\sigma,\theta)}{\bq^{A_1^{(i)}}_i(\sigma^\prime,\theta^\prime)}
\nonumber \\
&&\hspace{5cm} =\delta_{i,n-j-1}\delta^{A_1^{(i)}}{}_{A_2^{(i)}}\delta^{n-1}(\sigma-\sigma^\prime)\delta^{n-1}(\theta-\theta^\prime),
\eeqa
and the others are $0$.  

Let us take a Q-structure Hamiltonian function $\Theta$ on 
a QP-manifold of degree $n$, $\calM$.
The Q-structure functional 
on $\Map(T[1]\Sigma_{n-1}, \calM)$
is defined as 
\beq
    \vThe=\int_{T[1]\Sigma_{n-1}} d^{n-1}\sigma d^{n-1}\theta\ \ev^* \Theta.
\eeq
This is of degree $2$ and a Poisson bivector.
Since $(\bbOmega, \vThe)$ define a QP structure of degree $1$, 
a Poisson structure is defined.

From the assumption that there is no background structure except 
for a closed $n+1$ form $H$,
the Q-structure Hamilton function $\Theta$ of degree $n+1$ is 
uniquely determined as
\beqa
    &&\ev^* \Theta= 
\sum_{i=0}^{\floor{\frac{n}{2}}}\Big( k_{i}^{A_1^{(i)}A_2^{(i)}}
          \veta_{A_2^{(i)}\,{i+1}}
\veta_{A_1^{(i)}\,{n-i}}
+
\frac{1}{(n+1)!}H^{A_1^{(0)}\cdots 
A_{n+1}^{(0)}}\veta_{A^{(0)}_1\,1}
\cdots\veta_{A^{(0)}_{n+1}\, 1}
\Big ),
\eeqa
where $\floor{k}$ is the floor function which gives the largest integer
less than or equal to $k$.
The master equation of a Q-structure, $\sbv{\Theta}{\Theta}=0$,
is satisfied if and only if $H$ is a closed. 
where $\Phi \in \Map(T[1]\Sigma_{n-1}, \calM)$.
The derived bracket with respect to $\vThe$ derives the 
correct canonical commutation relations:
\beqa
    &&
\sbv{q^{A^{(i)}_1}_i(\sigma,\theta)}
{q^{A^{(j)}_2}_{j}(\sigma^\prime,\theta^\prime)}_{P.B.}
= \hatj^* 
\sbv{\sbv{\bq_i^{A^{(i)}_1}(\sigma,\theta)}{\vThe}}
{\bq_{j}^{A_2^{(j)}}(\sigma^\prime,\theta^\prime)},
\nom\\
    &&
\sbv{q^{A^{(0)}_1}_{n-1}(\sigma,\theta)}
{q^{A^{(0)}_2}_{n-1}(\sigma^\prime,\theta^\prime)}_{P.B.}
= \hatj^*
\sbv{\sbv{\bq_{n-1}^{A^{(0)}_1}(\sigma,\theta)}
{\vThe}}{\bq_{n-1}^{A_2^{(0)}}(\sigma^\prime,\theta^\prime)},
\eeqa
if the $\theta^{i+1}$ part of the superfields
$\veta^{A^{(i)}_1}_{i+1}\equiv (-1)^{i(n-i)}k_i^{A^{(i)}_1A^{(i)}_2}
\veta_{A^{(i)}_2\, i+1}$
is gauge fixed to 
$d_{\sigma} q^{A^{(i)}_1}_i$ 
and other auxiliary fields 
are projected by the pullback $\hatj^*$
after the calculation.

For example, the commutation relation of two terms
$f_{A^{(i)}_1}(x) d_{\sigma} q^{A^{(i)}_1}_i$ 
and
$g_{A^{(j)}_2}(x(\sigma^{\prime}))q^{A^{(j)}_2}_j$
on the original symplectic space is 
\beqa
    &&\hspace{0cm}\sbv{f_{A^{(i)}_1}(x(\sigma)) d_{\sigma} q^{A^{(i)}_1}_i(\sigma)}{g_{A^{(j)}_2}(x(\sigma^{\prime}))q^{A^{(j)}_2}_j(\sigma^\prime)}_{P.B.}\nom\\
    &&\hspace{0.5cm}=-\left((-1)^i\delta_{i,n-j-1}k_i^{A_1^{(i)}A_2^{(i)}}g_{A^{(j)}_2}df_{A^{(i)}_1} 
                      -\delta_{i,0}\delta_{j,n-1}k_0^{A^{(0)}A_2^{(0)}}
                      \frac{\del f_{A^{(i)}_1}}{\del x^{A^{(0)}}}g_{A^{(j)}_2}(x)d x_0^{A_1^{(0)}}\right)\delta^{n-1}(\sigma-\sigma^\prime)\nom\\ 
    &&\hspace{1cm}+\left((-1)^ik_i^{A_1^{(i)}A_2^{(i)}}\delta_{i,n-j-1}f_{A^{(i)}_1}g_{A^{(j)}_2}\right)(\sigma^\prime)
    d\delta^{n-1}(\sigma-\sigma^\prime),
\label{nDcommutation}
\eeqa
for $i\le j\ (i\ne n-1)$,
where $x^{A^{(0)}} = q^{A^{(0)}}_0$ is a canonical quantity 
of mass dimension zero.
Since 
$f_{A^{(i)}_1}(x) d_{\sigma} q^{A^{(i)}_1}_i$ 
and
$g_{A^{(j)}_2}(x(\sigma^{\prime}))q^{A^{(j)}_2}_j$
are contained in superfields
$f_{A^{(i)}_1}(\bx(\sigma,\theta)) \veta^{A^{(i)}_1}_{i+1}(\sigma,\theta)$
and 
$g_{A^{(j)}_2}(\bx(\sigma^{\prime},\theta^\prime))\bq^{A^{(j)}_2}_j(\sigma^{\prime},\theta^\prime)$,
(\ref{nDcommutation}) are calculated 
by the derived bracket on the QP manifold:
\beqa
&&\hspace{-1cm}\sbv{\sbv{f_{A^{(i)}_1}(\bx(\sigma,\theta)) \veta^{A^{(i)}_1}_{i+1}(\sigma,\theta)}{\vThe}}{g_{A^{(j)}_2}(\bx(\sigma^{\prime},\theta^\prime))\bq^{A^{(j)}_2}_j(\sigma^{\prime},\theta^\prime)}
\nom\\
&&\hspace{1.5cm}= -\left((-1)^i\delta_{i,n-j-1} k_i^{A^{(i)}_1A^{(i)}_2}g_{A^{(j)}_2}\frac{\del f_{A^{(i)}_1}}{\del \bx^{A^{(0)}}} \veta^{A^{(0)}}_{1}\right.\nom\\
&&\hspace{3cm}\left.-\delta_{i,0}\delta_{j,n-1}k_0^{A^{(0)}A_2^{(0)}}
                     \frac{\del f_{A^{(i)}_1}}{\del x^{A^{(0)}}}
g_{A^{(j)}_2}\veta^{A_1^{(0)}}_1
                     \right)\delta^{n-1}(\sigma-\sigma^\prime) \delta^{n-1}(\theta-\theta^\prime),
\eeqa
for $i\le j\ (i\ne n-1)$, which derives the correct first term in (\ref{nDcommutation}).
The commutation relation with respect to the graded Poisson bracket
\beqa
&&\sbv{f_{A^{(i)}_1}(\bx(\sigma,\theta)) \veta^{A^{(i)}_1}_{i+1}(\sigma,\theta)}{g_{A^{(j)}_2}(\bx(\sigma^{\prime},\theta^\prime))\bq^{A^{(j)}_2}_j(\sigma^{\prime},\theta^\prime)}\nom\\
&&\hspace{3cm}=(-1)^n f_{A^{(i)}_1}g_{A^{(i)}_2} k_i^{A^{(i)}_1A^{(i)}_2}\delta_{i,n-j-1}
\delta^{n-1}(\sigma-\sigma^\prime)\delta^{n-1}(\theta-\theta^\prime),
\eeqa
coincides with the correct coefficient of the second anomaly 
term in (\ref{nDcommutation})
if the systematic factor $(-1)^{i-n}$ is multiplied.
We can find that the correct terms are calculated
for more general terms.
For example, for 
$f_{A^{(i)}_1\cdots A^{(j)}_l}(x) d_{\sigma} q^{A^{(i)}_1}_i
\wedge \cdots \wedge d_{\sigma} q^{A^{(j)}_l}_j$,
the graded Poisson brackets 
for a superfields
$f_{A^{(i)}_1 \cdots A^{(j)}_l}(\bx) \veta^{A^{(i)}_1}_{i+1}
\cdots \veta^{A^{(j)}_l}_{j+1}$ 
derives the correct commutation relations after 
the pullback $\hatj^* \veta^{(i+1)A^{(i)}}_{i+1} 
= d_{\sigma} q^{A^{(i)}}_i$.
\medskip\\
\indent
Discussions in this section are summarized as follows.
Let 
$J_{k(\bJ_k)}(q^{A^{(i)}}_i(\sigma), d_{\sigma} q^{A^{(i)}}_{i}(\sigma))$
be a current of mass dimension $k$. 
The space of canonical quantities is extended to a graded manifold,
which is a QP manifold of degree $n$.
Then this current is mapped to a super function  
$\hatj :
J_{k(\bJ_k)}(q^{A^{(i)}}_i(\sigma), d_{\sigma} q^{A^{(i)}}_{i}(\sigma))
\longmapsto
\bJ_k(\bq^{A^{(i)}}_i(\sigma,\theta), \veta^{A^{(i)}}_{i}(\sigma,\theta))$
of degree $k$.
The original commutation relation is calculated as
\beqa
    &&\sbv{J_{k(\bJ_k)}(\sigma)}{J_{l(\bJ_l)}(\sigma^{\prime})}_{P.B.} 
= - J_{k+l+1-n (\courant{\bJ_k}{\bJ_l})}
\delta^{n-1}(\sigma - \sigma^\prime)
\nom\\  &&
\hskip5cm
+
\langle J_k, J_k \rangle
(\sigma^\prime)
d_{\sigma} \delta^{n-1}(\sigma-\sigma^\prime),
\eeqa
where
\beqa
\courant{\bJ_k}{\bJ_l} 
=  \sbv{\sbv{\bJ_k(\sigma)}{\Theta}}{\bJ_l(\sigma^{\prime})},
\eeqa
and $\langle J_k, J_l \rangle$ is obtained as
\beqa
\hatj^* \sbv{\bJ_k(\sigma,\theta)}{\bJ_l(\sigma^{\prime},\theta^\prime)}
= (-1)^{k-n} \langle J_k, J_l \rangle (\sigma^\prime)
\delta^{n-1}(\sigma-\sigma^\prime)\delta^{n-1}(\theta-\theta^\prime).
\eeqa
We have obtained the following theorem.
\begin{theorem}
A generalized current algebra in $n$ dimensions
has a structures of a QP manifold of degree $n$.
The anomalies cancel if and only if
current super functions are commutative under the Q-structure.
\end{theorem}
Let $\calL$ be a maximal subalgebra of $C^{\infty}(\calM)$,
such that elements are commutative $\sbv{\bJ_k}{\bJ_l}=0$
under the graded Poisson bracket 
and closed under the derived bracket 
$\sbv{\sbv{\bJ_k}{\Theta}}{\bJ_l}\in \calL$.
The anomaly cancellation condition is that 
a set of current functions is 
$\calL$, which is a generalization of the Dirac structure.
In fact, if a target space is a QP manifold $T^*[n]E[1]$, 
then a QP structure defines a generalized Courant-Dorfman bracket 
on $\Gamma E\oplus\wedge^{n-1}E^{*}$
from Example \ref{liealoid}.
In this example, 
the anomaly cancellation condition is equivalent to 
the restriction of a generalized Dirac structure $L$
with respect to the generalized Courant-Dorfman bracket.

Since currents $J_{n-1}$ of dimensions $n-1$ correspond
to super functions $\bJ_{n-1}$ of degree $n-1$, 
the $J_{n-1}$ part of current algebras has 
a structure of the Loday algebroid from the theorem \ref{QPloday}.

\section{Conclusions and Discussion}
\noindent
We have investigated generalized current algebras in any dimension.
Symplectic structures of canonical conjugates have been
reformulated by a QP manifold.
Current algebras and anomalies in $n$ dimensions have 
structures of an algebroid characterized by a QP manifold
of degree $n$. 
The anomaly cancellation conditions are 
equivalent to the condition 
that current functions on a QP manifold consist of a commutative
subalgebra.

A QP manifold structure of degree $n$ 
of current algebras in $n$ dimensions
suggests a holographic correspondence of
them with a quantum field theory in $n+1$ dimensions.
Because a topological field theory in $n+1$ dimensions 
has a structure of a QP manifold of degree $n$
via the AKSZ construction.
These will be a generalization of the correspondence of the WZW model 
in two dimensions to
the Chern-Simons gauge theory in three dimensions.

More generalizations of current algebras have been analyzed 
in two dimensions in \cite{Ekstrand:2009qz}.
Our discussion can be extended to that case and 
these current algebras will be reconstructed 
in terms of a QP manifold and generalized to higher dimensions.
Our results will be generalized to a manifold with boundaries.
This case is connected to membrane theories, such as D-branes
and M-branes.

\section*{Acknowledgments}

The author (N.I.) would like to thank K.Uchino 
for valuable discussions and comments.
This work is supported by Maskawa Institute, Kyoto Sangyo University.

\appendix
\section{Appendix}
\subsection{Current Algebras in Three Dimensions}
\noindent
Current algebras in Eqs.~(\ref{3dcurrentalgebra}) and
(\ref{3dalgebra}) calculated as components
are following:
\beqa
    &&\{J_{0(f)}({\sigma}),J_{0(f^\prime)}({\sigma^\prime})\}_{P.B.}=0,\nom\\ 
 \noalign{\vskip 2mm}
    &&\{J_{1i(\alpha,u)}({\sigma}),J_{0(f^\prime)}({\sigma^\prime})\}_{P.B.}=0,\nom\\
 \noalign{\vskip 2mm}
    &&\{J_{2ij(G,K,F,H,E)}({\sigma}),J_{0(f^\prime)}({\sigma^\prime})\}_{P.B.}
    =-\epsilon_{ij}G^{I}(x(\sigma))\frac{\del f^\prime(x(\sigma))}{\del x^I}
\delta^2(\sigma-\sigma^\prime),\nom\\
 \noalign{\vskip 2mm}
    &&\{J_{1i(\alpha,u)}({\sigma}), J_{1j(\alpha^\prime,u^\prime)}({\sigma}^\prime)\}_{P.B.}
    =\epsilon_{ij} u_{A}(x({\sigma}))u^\prime_{B}(x({\sigma})) k^{AB}\delta^2({\sigma}-{\sigma}^\prime),\nom
\eeqa
\beqa
    &&\{J_{2ij(G,K,F,B,E)}({\sigma}),J_{1k(\alpha^\prime,u^\prime)}({\sigma^\prime})\}_{P.B.}\nom\\ 
    &&\hspace{1cm}=-\epsilon_{ij}J_{1k(\bar\alpha,\bar u)}\delta^2({\sigma}-{\sigma}^\prime)
    +\epsilon_{ij}(-k^{AB}u_A^\prime K_B+\alpha_I^\prime G^{I})(x({\sigma}^\prime))
    \partial_k \delta^2({\sigma}-{\sigma}^\prime),\nom\\
  \noalign{\vskip 2mm}     
    &&\{J_{2ij(G,K,F,B,E)}({\sigma}),J_{2kl(G^\prime,K^\prime,F^\prime,B^\prime,E^\prime)}({\sigma}^\prime)\}_{P.B.}\nom\\
    &&\hspace{1cm}=-\epsilon_{ij}\epsilon_{kl}J_{2(\bar G,\bar K,\bar F,\bar B,\bar E)}\delta^2({\sigma}-{\sigma}^\prime)\nom\\
    &&\hspace{1.5cm}-\epsilon_{ij}\epsilon_{kl}
\left(G^{\prime I}B_{IJ}+G^{I}B_{IJ}^\prime
            +k^{AB}(E_{AJ}K^\prime_B+E^\prime_{AJ}K_B)\right)
                 \epsilon^{mn}(\partial_m x^J\partial_n)
\delta^{2}({\sigma}-{\sigma}^\prime)\nom\\
    &&\hspace{1.5cm}-\epsilon_{ij}\epsilon_{kl}
    \left(G^{\prime I}E_{AI}+G^{I}E^\prime_{AI}
              +k^{BC}(K_CF_{AB}^\prime+K_C^\prime F_{AB})\right)
              \epsilon^{mn}(q^A_{m}\partial_n)
\delta^{2}({\sigma}-{\sigma}^\prime),\qquad\quad
\eeqa
where
\beqa
&&    \bar\alpha_{I}=G^{J}\frac{\partial \alpha_{I}^\prime}{\partial x^J}
                 +\alpha_{J}^\prime\frac{\partial G^{J}}{\partial x^I}
                 +k^{AB}\left(-u_{A}^\prime\frac{\partial K_B}{\partial x^I}
                              +u_{A}^\prime E_{BI}\right),\nom 
\\
&&    \bar u_{A}=G^{I}\frac{\partial u_{A}^\prime}{\partial x^I}
-k^{BC}u_{B}^\prime F_{CA},\nom
\eeqa
and
\beqa
   {\bar G}^I&=&G^J\frac{\partial G^{\prime I}}{\partial x^J}
                -G^{\prime J}\frac{\partial G^{I}}{\partial x^J},\nom
\\
 \noalign{\vskip 2mm}
   {\bar K}_A&=&G^I\frac{\partial K_A^\prime}{\partial x^I}
               -G^{\prime I}\frac{\partial K_A}{\partial x^I}
                +k^{BC}K_B^\prime F_{AC}+E_{AI}G^{\prime I},\nom\\
 \noalign{\vskip 2mm}
   {\bar F}_{AB}&=&G^I\frac{\partial F^{\prime}_{AB}}{\partial x^I}
                  -G^{\prime I}\frac{\partial F_{AB}}{\partial x^I}
                  +k^{CD}(F_{AC}F^\prime_{DB}-F_{BC}F^\prime_{DA}),\nom\eeqa
\beqa
   \bar B_{IJ}&=&G^K\frac{\partial B_{IJ}^\prime}{\partial x^K}
                 -\frac{\partial G^K}{\partial x^I}B^\prime_{JK}
                 -\frac{\partial G^K}{\partial x^J}B^\prime_{KI}
                 -G^{\prime K}\left(\frac{\partial B_{IJ}}{\partial x^K}
                 +\frac{\partial B_{JK}}{\partial x^I}
                 +\frac{\partial B_{KI}}{\partial x^J}\right)+G^KG^{\prime L}H_{KLIJ}\nom\\
       &&\hspace{-1.5cm}+k^{AB}\left((E_{AJ}E^\prime_{BI}-E_{AI}E^\prime_{BJ})
                 +K_B^\prime\left(\frac{\partial E_{AJ}}{\partial x^I}
                                  -\frac{\partial E_{AI}}{\partial x^J}\right)
                 +\left(E_{BJ}^\prime\frac{\partial K_A}{\partial x^I}
                 -E_{BI}^\prime\frac{\partial K_{A}}{\partial x^J}\right)\right),\nom\\
 \noalign{\vskip 2mm}
    \bar E_{AI}&=&
        G^J\frac{\partial E_{AI}^\prime}{\partial x^J}+\frac{\partial G^J}{\partial x^I}E^\prime_{AJ}
       +G^{\prime J}\left(\frac{\partial E_{AJ}}{\partial x^I}
                          -\frac{\partial E_{AI}}{\partial x^J}\right)\nom\\
       &&\hspace{2cm}+k^{BC}\left(E_{BI}F^\prime_{CA}-F_{BA}E^\prime_{CI}           
         +\frac{\partial F_{AB}}{\partial x^I}K_C^\prime
         -\frac{\partial K_B}{\partial x^I}F_{CA}^\prime\right).
\eeqa
Anomaly cancellation conditions are given by
\[
   -k^{AB}u_A^\prime K_B+\alpha_I^\prime G^{I}=0,
\]
\[
    G^{\prime I}B_{IJ}+G^{I}B_{IJ}^\prime
            +k^{AB}(E_{AJ}K^\prime_B+E^\prime_{AJ}K_B)=0,
\]
\beq
    G^{\prime I}E_{AI}+G^{I}E^\prime_{AI}
              +k^{BC}(K_CF_{AB}^\prime+K_C^\prime F_{AB})=0.
\eeq



\end{document}